\documentclass{aa}
\usepackage{natbib}
\bibpunct{(}{)}{;}{a}{}{,}
\usepackage{amssymb}
\usepackage{graphicx}
\begin{document}
\title{Relativistic emission lines from accreting black holes}
\subtitle{The effect of disk truncation on line profiles}
\author{Andreas M\"uller \& Max Camenzind}
\titlerunning{Relativistic emission lines from accreting black holes}
\authorrunning{A. M\"uller \& M. Camenzind}
\institute{Landessternwarte Koenigstuhl, D--69117 Heidelberg,Germany}
\offprints{ Andreas M\"uller,
\email{amueller@lsw.uni-heidelberg.de}}
\date{Received 30 July 2003 / Accepted 22 September 2003}
%
%
%
%
%
\abstract{Relativistic emission lines generated by thin accretion disks
around rotating black holes are an important diagnostic tool
for testing gravity near the horizon. The iron K--line is of
special importance for the interpretation of the X--ray emission
of Seyfert galaxies, quasars and galactic X--ray binary systems.
A generalized kinematic model is presented which includes radial drifts
and non--Keplerian rotations for the line emitters. The resulting
line profiles are obtained with an object--oriented ray tracer operating
in the curved Kerr background metric. The general form of the
Doppler factor is presented which includes all kinds of poloidal
and toroidal motions near the horizon. The parameters of the model
include the spin parameter, the inclination, the truncation and outer
radius of the disk, velocity profiles for rotation and radial drift,
the emissivity profile and a multi--species line--system. \\
The red wing flux is generally reduced when radial drift is included as
compared to the pure Keplerian velocity field.
All resulting emission line profiles can be classified as triangular,
double--horned, double--peaked, bumpy and shoulder--like. Of
particular interest are emission line profiles generated by truncated
standard accretion disks (TSD). It is also shown that the emissivity law
has a great influence on the profiles. The characteristic shoulder--like
line profile observed for the Seyfert galaxy MCG--6--30--15 can
be reproduced for suitable parameters.
}
\maketitle
%
%
%
%
\section{Introduction}
Emission lines originating from regions where the influence of curved space--time can not be neglected are found in
accreting black hole systems like Active Galactic Nuclei (AGN), microquasars and Galactic Black Hole Candidates (GBHC).
These astrophysical systems exhibit hot plasma falling into the deep potential well of black holes. The ion species in the plasma
still have some electrons on lower shells so that there are several transitions between electron states that generate emission lines
in the X--ray range. The most prominent line is the Fe K$\alpha$ line at 6.4 keV rest frame energy (near-neutral iron). Additionally, the Fe K$\beta$ line
at 7.06 keV and rarely Ni K$\alpha$ at 7.48 keV and Cr K$\alpha$ at 5.41 keV can be applied to fit X--ray emission line systems. The fluorescence process and its
transition energies depend on the ionization state of the species. One can distinguish hot and cold regimes of line emission which can be
described by an ionization parameter \citep{BRF}. Generally, X-ray astronomers observe a multi--component emission line complex
of these species. The Fe K$\alpha$ line dominates the line system due to its high relative strength and is often indicated in unreduced
global X-ray spectra. \\
Broad X--ray emission lines originate from the hot inner parts of accreting black hole systems, typically from radii comparable to the
radius of marginal stability. Naturally, these lines hint for a proximity of hot and cold material: hot material can be found in an
optically thin corona, a place where cold seed photons are Comptonized to hard radiation. It is expected that the corona becomes optically
thick at super-Eddington accretion rates and is dominated by Comptonization. This latter scenario seems to hold for narrow-line Seyfert-1
galaxies (NLS1s), whereas classical broad-line Seyfert-1 galaxies (BLS1s) accrete at sub-Eddington to Eddington rates.\\
Relatively cold material hides in the radiatively--efficient, geometrically thin and
optically thick accretion disk, the so called {\em standard disk}, hereafter SSD ({\em Shakura--Sunyaev Disk}) \citep{SS}. The resulting
structure equations for this standard disk have been generalized to the relativistic case later \citep{NT}. The hard coronal input radiation
usually described by a power law illuminates the cold disk. As a consequence, the surface of the disk is ionized in a thin layer. In a fluorescence
mechanism the hard radiative input is absorbed (threshold at 7.1 keV) and re--emitted in softer fluorescence photons. But in the dominant competitive
process, the auger effect, the excited ions emit auger electrons and enrich the environment with hot electrons. One of the first
applications of this X--ray fluorescence was performed at the microquasar Cyg X--1 \citep{fab3}. These observations were supplemented
by radio--quiet AGN, e.g. Seyfert galaxies, showing the same X--ray reflection signatures \citep{poun,tan,RN}. \\
But there are also sources exhibiting narrow X--ray emission lines. The narrow lines do not show distortion by relativistic effects.
It has been proposed that narrow X--ray emission lines can be explained by reflection at the large--scale dusty
torus \citep{turn}. Here, the narrow line is overimposed on broad line profiles. This interpretation has been applied to some sources,
like Seyfert galaxies, e.g. MCG--6--30--15 \citep{wilms}, MCG--5--23--16 \citep{DGS} and Quasars, e.g. Mrk 205 \citep{boll}.\\
The absence of the line feature in several Sefert--1 galaxies \citep{pfeff} can be explained by global
accretion models. The first possible interpretation is that the inner edge of the disk seems to be far away from the black hole.
If the fluorescence process is still possible in that distance, the skewness of the X--ray emission line vanishes and becomes rather
Newtonian-like. In even larger distances, it is supposed that the emission line feature disappears completely because irradiation and
fluorescence is reduced. Accretion theory states that this scenario holds for {\em low}
accretion rate: then the transition radius between standard accretion disk and optically thin disk is large \citep{esin}. The mass
determinations for Seyfert black holes confirm this interpretation while being in the intermediate
range of $10^{6}$ to $10^{8}$ $\mathrm{M}_{\odot}$. These lighter supermassive black holes seem to accrete very weakly. \\
X--ray spectra of AGN and X--ray binaries are dominated by broad Comptonized continua. The properties of these spectra are discussed in
the next section. The first step for astronomers is to reduce the observed data files and to extract the line profile by subtracting this
global continuum emission. This procedure is already complicated because the intrinsic curvature of the Comptonized radiation is still
topic of an ongoing debate. \\
\begin{figure}
  \rotatebox{0}{\includegraphics[height=6.62cm,width=9cm]{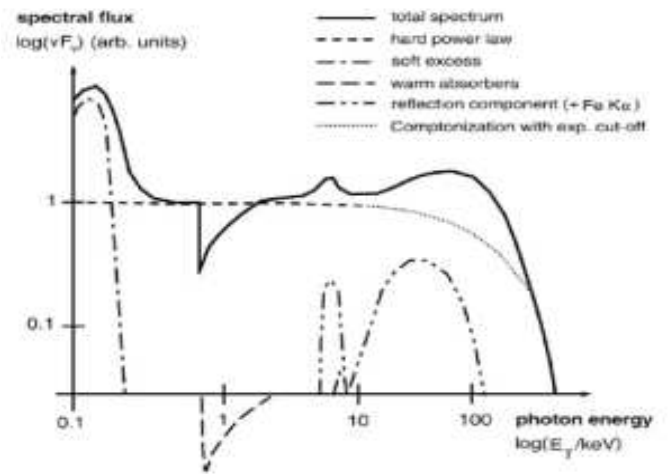}}
  \caption{Prototype X--ray spectrum of a type--1 AGN (illustration idea taken from Fabian, 1998). The
  spectrum is dominated by the global {\em Comptonized continuum}. At low energies, there is the {\em soft excess},
  in principle a composition of some black bodies belonging to the disk divided into rings. At energies around 1 keV,
  one can recognize a complex of absorption dips, originating from the {\em warm absorbers}. The {\em reflection component}
  consists of a broad bump peaking around 20 keV and the prominent fluorescence emission line complex around 6.4 keV.} \label{fig:xspec}
\end{figure}
The resulting emission line profile serves for diagnostics and to constrain essential parameters of the accreting black hole system.
So, astronomers found an X--ray imprint of black holes and can derive properties concerning the relative position from
the inner accretion disk ({\em standard disk}) to the observer, the \textbf{inclination $i$} or the rotational state of the black hole,
the \textbf{Kerr rotational parameter $a$}, the extension of the geometrically thin and optically thick standard disk, like \textbf{inner radius
$r_\mathrm{in}$} and \textbf{outer radius $r_\mathrm{out}$}, the \textbf{velocity field of the plasma}, separated in toroidal $v^{\Phi}$, poloidal
$v^{\Theta}$ and radial $v^\mathrm{r}$ components and the relative position of hot corona to cold accretion disk.\\
Let us follow up the theoretical steps from black hole theory needed here until one gets simulated emission line profiles:
Sect. \ref{sec:raytr} introduces the metric of rotating black holes, the Kerr geometry, and provides the standard co--ordinate
system in {\em Boyer--Lindquist form}. Then, the \textbf{generalized GR Doppler factor} is presented. The derivation of this quantity and
its signification in GR ray tracing is elaborated, too. \\
In Sect. \ref{sec:emit}, realistic plasma kinematics resulting from hydrodynamic and magnetohydrodynamic accretion theory is
discussed. The Kerr geometry has special features that are valid independently from the accretion model! This is accessable by
analytical investigations. \\
Another basic ingredient to calculate emission lines and spectra in general is the \textbf{emissivity law} that determines the
emission behaviour of the line emitting region. In Sect. \ref{sec:emiss}, alternative emissivity shapes besides
th classical single power law are applied.\\
Ray racing provides in a first step disk images. It may be possible to resolve these images observationally in near future. In any
case, disk images as presented in Sect. \ref{sec:visdisk} are very instructive unrevealing GR effects and influence of plasma dynamics.\\
In Sect. \ref{sec:feline}, the calculation and basic properties of the emission lines are discussed. \\
In the final Sect. \ref{sec:linzoo}, the parameter space of relativistic emission lines is investigated and suitable criteria
to fix line features are presented. One can apply these criteria to observed line profiles and derive a classification in the line jungle.
%
%
%
%
\section{X--ray spectra and topology of type--1 AGN} \label{sec:xrayspec}
The data quality increased conspicuously with the launch of the space--based X--ray observatories Chandra and XMM--Newton. The
challenge in interpreting X--ray spectra of AGN and XRBs is to correctly reduce the data and subtract the continuum.
Afterwards one can discuss the formation of other spectral components, such as emission lines.\\
X--ray spectra of {\em Seyfert galaxies} have been investigated for several years. These types of AGN were the first objects that gave
evidence for rotating black holes \citep{tan} and confirmed the paradigm of the AGN engine: an accreting supermassive black hole producing 
high multi-wavelength luminosity. Only a few years later, high--resolution observatories allowed the
extraction of iron K emission lines in more distant objects, Quasars \citep{yaq}. In the course of the \textbf{unification scheme of AGN}, the
different AGN types are considered as different evolution phases of a prototype AGN with some pecularities that mainly arise by differencies
in orientation of the accretion disk, accretion rate and black hole mass. Especially, the discrimination between Seyferts and Quasars has become
somewhat arbitrary. Only the amount of radiation, $\approx 10^{11}$ $L_{\odot}$ for Seyferts and $10^{13}$ to $10^{15}$ $L_{\odot}$ for Quasars, is
the main difference between these two subclasses. Concerning the iron emission line feature, it can be stated that Quasars exhibit increased ionization
of the standard disk surface due to enhanced central luminosity. Therefore, weaker emission line signatures are expected \citep{RN}.  \\
The dichotomy phenomenon caused by disk orientation and dust torus obscuring, a generalization that can be applied
to Seyferts {\em and} Quasars, is well--known: one can distinguish \textbf{type--1} (low inclination) and \textbf{type--2 AGN} (high inclination).
Hence, it is for observational reasons only possible to study relativistic emission lines of AGN of type 1.\\
\begin{figure}
  \rotatebox{0}{\includegraphics[height=4.8cm,width=9cm]{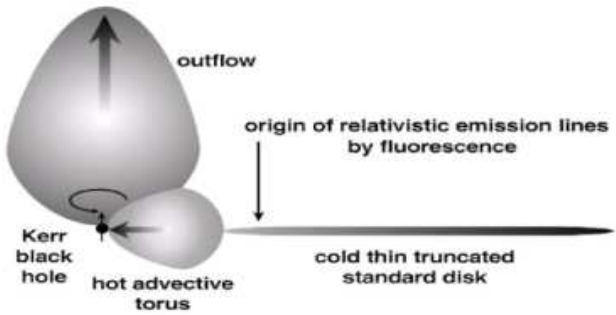}}
  \caption{Illustration of topological elements in the innermost region of accreting black hole systems. The size and morphology of these elements
  depend on accretion rate, black hole mass and radiative cooling.} \label{fig:topo}
\end{figure}
X--ray spectra of Seyfert galaxies in the range from 0.1 to several 100 keV show many features as illustrated in Fig. \ref{fig:xspec}:
overlaying the complete X--ray range there is a \textbf{continuum} originating from Comptonized radiation. This is the direct radiation that
reaches the observer coming from a hot corona ($T\approx$ 30 keV). The photon number flux per unit energy is usually modelled with a power
law distribution, $F_\mathrm{N}(E)\propto E^{-\Gamma}$. The soft input radiation from the cold accretion disk is
reprocessed via unsaturated inverse Compton scattering by ultra--relativistic electrons in the corona. Fig. \ref{fig:topo} depicts in a
simple way one probable geometry of an accreting black hole systems that hold especially for AGN and microquasars, known as "sphere+disk
geometry". Matter has angular momentum and therefore spirales down in a flat and cold standard disk. The argument for the
flatness is because of vertical collapse due to efficient radiation cooling. The relative scale heigth $H/r$ of these standard disks
is only $10^{-3}$ \citep{SS}. The matter within this cold disk emits a multi-color blackbody
spectrum. This is simply the superposition of a sequence of Planck spectra, e.g. rings each with a temperature $T_\mathrm{i}$. The innermost ring
is the hottest one, $T_\mathrm{in}\geq T_\mathrm{i}\geq T_\mathrm{out}$. The accretion rate determines the transition radius and this one fixes the
temperature of the inner edge of the disk. At high accretion rates, the inner edge is significantly hot so that a soft excess can be observed around
1 keV. At small accretion rates the soft excess lacks \citep{esin}. The cold and thin accretion disk delivers the soft seed photons that are Comptonized
in the hot, optically thin corona. The topology of the corona is still one of the essential open questions in X--ray spectroscopy. Different
assumptions such as slab and patchy corona models have been made \citep{RN}. The observed hard spectra suggest rather "sphere+disk geometries" because
slab and patchy corona models are more efficiently radiative--cooled by seed photons from the cold disk. The reverberation mapping technique may enlighten
the spatial position of corona to accretion disk. Theoretically, it will be task of upcoming relativistic MHD simulations to unreveal the corona--disk
geometry depending on accretion rate and radiative cooling. Apparently, it turns out that the coronal emissivity is significantly enhanced at the inner edge of the standard disk \citep{MF} and
that MHD dissipation plays a crucial role to invalidate the zero--torque boundary condition. Then, magnetically--induced torques at radii comparable to
the radius of marginal stability are expected to increase dissipation. This mechanism may provide the higher disk emissivities $\epsilon(r)\propto r^{-4.5}$
as recently observed in Seyfert galaxy MCG--6--30--15 \citep{wilms}. \\
Here it is assumed that the corona is almost identical with an advection--dominated torus that forms by feeding from the standard disk. An alternative may
be the quasi--spherical rather Bondi--accreting {\em Advection--Dominated Accretion Flow} (ADAF) \citep{naryi} that forms under other circumstances in accretion
theory, e.g. depending on the accretion rate. The transition region where the standard disk adjoins the hot torus is unstable. Hot corona and cold disk
may constitute a sandwich configuration where the inner edge of the cold disk oscillates in radial direction \citep{josep}. This ADAF--SSD transition region seems to
depend on the efficiency of different cooling channels as can be found in hydrodynamic simulations \citep{man,jose}.
More recent pseudo--Newtonian hydrodynamic simulations incorporating radiation cooling via synchrotron radiation, bremsstrahlung and
Comptonization with the consideration of conduction in a two--component plasma suggest \textbf{disk truncation} \citep{huj}. This means
that the accretion disk can cut off even at radii {\em larger} than the radius of marginal stability, $r_\mathrm{ms}$. \textbf{Truncation of standard disks
(TSD)} in general softens the argument for observed evidences of rapidly spinning black holes. This is because truncated disks do not extend to
$r_\mathrm{ms}$ inwardly. However the decreasing $r_\mathrm{ms}$ for faster spinning black holes was always the argument for evidence of Kerr black
holes \citep{iwa}. \\
Particularly, the cooling time--scale of synchrotron radiation is typically in the millisecond domain and cools very fast the hot accretion flow. The
{\em Truncated Disk -- Advective Torus} (TDAT) scenario \citep{huj} produces a hot inner torus that is not stable either. It is destablized by accretion and
matter free--falls into the black hole. But a small fraction of matter can escape the black hole and is transported along magnetic field lines in polar
regions. This drives outflows that consist of disk winds on one hand and of ergospheric Poynting fluxes on the other hand. This outflow is supposed to
feed the large--scale jets of AGN or the blobs of microquasars. In a unified view, the total accretor mass, $M$, and the total mass accretion rate,
$\dot M$, are strongly favoured to determine the spectral state and the jet injection rate and the continuity of the outflow. Relativistic emission
lines may serve as a diagnostic tool to fix these parameters. \\
The continuum of the hot corona reveals an intrinsical curvature and cuts off at the high--energy branch at several hundreds of keV. The cut--off at
$3k_\mathrm{B}T_\mathrm{C}$ indicates the plasma temperature in the corona \citep{RL}. At the low energy branch, some sources show a \textbf{soft excess} with a bumpy profile.
This can be identified with a multi-color blackbody originating from the cold Compton--thick accretion disk. At ~1.0 keV there is a complex structure of
\textbf{warm absorbers}, typically a combination of adjacent absorption edges of a variety of species (\ion{C}{V}, \ion{O}{VII}, \ion{O}{VIII},
\ion{Ne}{IX} etc.). These dips and their potentially relativistic broadening are still topic of an ongoing debate \citep{mas,lee}.
The \textbf{reflection component} consists of a system of emission lines, most prominently the iron K$\alpha$ at 6.4 keV typically, and a
\textbf{broad bump} peaking at 20 to 30 keV. The reflection bump is a consequence of hard radiation from the corona hitting the cold accretion disk
that is then reflected to the observer \citep{reyn}. The origin of the iron K$\alpha$ line is well understood: at typical temperatures of $10^{5}$ to
$10^{7}$ keV in the inner accretion disk, iron is ionized but not completely stripped; the K-- and L--shell are still populated. First, iron is excitied
by photo--electric absorption. The threshold lies at $\approx$ 7.1 keV so that only hard X--ray photons from the corona can release this primary process.
Then, there are two competing processes: the dominant process (66 \% probability) is the Auger effect where the excitation energy is emitted with an
Auger electron. This mechanism is non--radiative and enriches the disk plasma with electrons. The second and here essential process is \textbf{fluorescence}
(34 \% probability): one electron of the L--shell undergoes the transition to the K--shell accompanied by the emission of the iron K$\alpha$
fluorescence photon with 6.4 keV rest frame energy\footnote{We neglect here any distinction of the K$\alpha_{1}$ line at 6.404 keV and K$\alpha_{2}$
line at 6.391 keV as can be found in atomic physics.}. The emission line energy and the existence of the fluorescence line in
general depend on the ionization state of the material \citep{RFY}. \\
The significance of other elements contributing to the reflection component is rather low, because the abundance, relative strength and fluorescence yield is low
compared to that of iron \citep{reyn}. At most, the iron K$\beta$ at 7.06 keV or nickel at 7.48 keV \citep{wang} and chromium at 5.41 keV could contribute
marginally. Besides, the accretion disk has only a thin ionization layer. The line producing layer has only the depth of 0.1 to 1\% of the disk
thickness \citep{matt2}. Therefore the contributions of ionized iron from deeper regions -- or even a radiation transfer problem from lower ionized
slabs -- can be neglected due to high optical depths.
%
%
%
%
\section{Ray tracing in the Kerr geometry} \label{sec:raytr}
\begin{figure}
  \rotatebox{0}{\includegraphics[height=5.09cm,width=9cm]{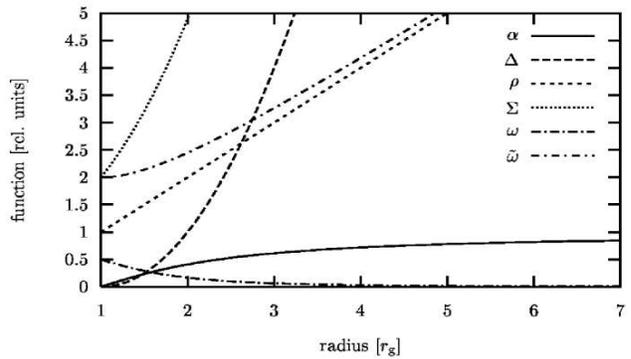}}
  \caption{Radial behaviour of the Boyer--Lindquist functions at extreme Kerr ($a=1.0$) in the equatorial plane. Relativistic
  units were used.} \label{fig:boy}
\end{figure}
\begin{figure}
  \rotatebox{0}{\includegraphics[height=4.82cm,width=9cm]{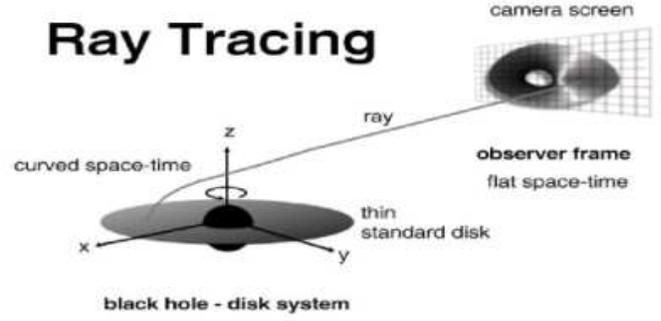}}
  \caption{Schematical representation of a Kerr ray tracer. The light rays start at the screen ({\em back tracking}) and hit the disk surface in the
  equatorial plane. On the screen the lensed image is formed as seen by a distant observer. Our solver folds {\em directly} from equatorial plane to screen.} \label{fig:rt}
\end{figure}

The Kerr metric describes rotating uncharged black holes. This vacuum solution of the Einstein field equations \citep{kerr} generalizes the static
Schwarzschild solution. The line element fullfills the standard form of stationary and axisymmetric space--times
\begin{equation}
ds^{2}=e^{2\Phi}dt^{2}-e^{2\Psi}(d\Phi-\omega dt)^{2}-e^{2\mu_\mathrm{r}}dr^{2}-e^{2\mu_{\Theta}}d\Theta^{2}.
\end{equation}
The Kerr geometry in Boyer--Lindquist form is determined by the line element
\begin{equation}
ds^{2}=-\alpha^{2}dt^{2}+\tilde\omega^{2}(d\Phi-\omega dt)^{2}+\left(\rho^{2}/\Delta\right)dr^{2}+\rho^{2} d\Theta^{2},
\end{equation}
with the canonical Boyer--Lindquist functions
\begin{eqnarray} \label{eq:boy}
\alpha & = & \frac{\rho\sqrt{\Delta}}{\Sigma}, \\
\Delta & = & r^{2}-2Mr+a^{2}, \\
\rho^{2} & = & r^{2}+a^{2}\cos^{2}\theta, \\
\Sigma^{2} & = & (r^{2}+a^{2})^{2}-a^{2}\Delta\sin^{2}\theta, \\
\tilde{\omega} & = & \frac{\Sigma}{\rho}\sin\theta, \\
\omega & = & \frac{2aMr}{\Sigma^{2}},
\end{eqnarray}
$M$ is the mass and $a$ the spin parameter of the black hole.\\
The Boyer--Lindquist co--ordinates are pseudo--spherical. The expressions above are called frame--dragging frequency or potential for angular
momentum $\omega$, cylindrical radius $\tilde{\omega}$, lapse function or redshift factor $\alpha$ and some geometrical functions $\Delta$ and $\rho$. \\
The rotation of the black hole can be parametrized by its specific angular momentum $a$, the so called \textbf{Kerr parameter}. Relativistic units
$G=M=c=1$ are used unless otherwise stated. Then, the natural length--scale is the {\em gravitational radius}, $r_\mathrm{g}=GM/c^{2}$, and $a$ varies
between $-1$ and $1$. The interval $a\in[-1,0]$ involves retrograde rotation between black hole and disk,
the interval $a\in[0,1]$ means prograde rotation. From black hole theory, it is generally possible that $a$ takes all values in this interval, but
accretion theory suggests that there is mainly one type of black holes: rotating Kerr black holes near its maximum rotation ($a\approx 1$).
This is because black holes of mass $M$ are spun up to $a\simeq\Delta m/M$ by accreting a mass $\Delta m$. At least the supermassive
black holes in AGN that accrete for long times are expected to rotate very rapidly, $a\geq 0.9$. It is possible to extract angular momentum from black
holes via Blandford-Znajek \citep{bz} and Penrose processes \citep{PF}. But these effects may not be efficient enough to slow down the black hole rotation significantly. \\
Fig. \ref{fig:boy} illustrates the radial dependence of the Boyer--Lindquist functions in the equatorial plane ($\theta=\pi/2$) at the maximum value of
black hole rotation, $a=1$. One can see the frame--dragging frequency $\omega$ that increases significantly at smaller radii and reaches
$\omega(r_\mathrm{H})=\Omega_\mathrm{H}=1/2$ in relativistic units at the horizon. This behaviour shows that dragging of inertial frames is
important when reaching the ergosphere. Another essential feature is the steep decrease of the redshift factor $\alpha$. This is mainly the cause
that emission is strongly suppressed when the distance becomes smaller than approximatly the radius of marginal stability, $r_\mathrm{ms}$,
which satisfies
\begin{eqnarray}
r_\mathrm{ms} & = & M\left(3+Z_{2}\mp\sqrt{(3-Z_{1})(3+Z_{1}+2Z_{2})}\right), \\
Z_{1} & = & 1+\left(1-\frac{a^{2}}{M^{2}}\right)^{1/3}\left(\left(1+\frac{a}{M}\right)^{1/3}+\left(1-\frac{a}{M}\right)^{1/3}\right) \nonumber\\
Z_{2} & = & \sqrt{3\frac{a^{2}}{M^{2}}+Z_{1}^{2}}\nonumber,
\end{eqnarray}
(upper sign prograde, lower sign retrograde orbits). \\
Our object--oriented code \textbf{KBHRT} ({\em Kerr Black Hole Ray Tracer}) \citep{andy} programmed in C++ is used to derive in the first step the
image of a relativistic disk and in the second step to calculate the line flux by integrating over this image and weighting with the relativistic
generalized Doppler factor and the radial disk emissivity profile. The imaging of the disk is not observationally possible because one
can not (yet) resolve these distant accreting black hole systems. But these images show nicely the influence of curved space--time on photon paths. The
flux is indeed the observationally accessible information and can be determined afterwards.\\
The code which is based on a C--code \citep{fant} calculates the photon path by integrating the GR geodesics equation using the complete set of integrals of
motion. Besides the canonical conservative quantities like mass, energy and angular momentum, the Kerr geometry exhibits a fourth one: the \textbf{Carter
constant} \citep{cart}. This quantity is associated with the radial and poloidal photon momentum. The second order null geodesics equation reduces by means
of these four conservatives to a set of four first order ordinary differential equations which can easily be integrated via elliptical integrals or
Runge--Kutta schemes. The code is very fast because the position on the disk (r, $\theta$, $\Phi$) is folded directly onto the observers screen (x, y)
{\em without} following the path stepwise between these two 2D objects (see Fig. \ref{fig:rt}). This is an alternative to the first historical method
using transfer functions \citep{cunn} which is still in use \citep{brom,spei} or direct shooting, a method that follows {\em the complete ray} from
locus of emission to observer.
The constraint of dealing only with flat disks that lay infinitely thin in the equatorial plane is a good approximation to the vertical collapsed
geometrically thin standard disks, however it is a restriction of our code. One recent issue shows the direction of future work, where the ray
tracing code has to be coupled to hydrodynamical or MHD accretion models \citep{AR}. This coupling allows to study the variable emission near
black holes. \\
The \textbf{KBHRT} code encapsulates all relevant tools to visualize relativistic disks and emission lines. One can study the distribution of the
generalized Doppler factor allover the disk or the emission itself. \\
The \textbf{generalized Doppler factor} is defined by
\begin{equation}
g\equiv\frac{\nu_\mathrm{obs}}{\nu_\mathrm{em}}=\frac{1}{1+z}=\frac{\hat{p}^\mathrm{t}_\mathrm{obs}}{\hat{p}^\mathrm{t}_\mathrm{em}},
\end{equation}
where $\nu_\mathrm{obs}$ and $\nu_\mathrm{em}$ correspond to observed and emitted photon momenta, $z$ to the redshift and $\hat{p}$ denotes the rest frame of the
plasma. The {\em Zero Angular Momentum Observer} (ZAMO) or {\em Bardeen observer} momenta have to be boosted into the rest frame of the plasma
to derive $g$ by means of a local Lorentz transformation,
\begin{eqnarray}
\hat{p^\mathrm{t}} & = & \gamma\left[p^\mathrm{(t)}-v^\mathrm{(j)}p_\mathrm{(j)}\right], \nonumber\\
\hat{p^\mathrm{i}} & = &
p^\mathrm{(i)}+\frac{1}{v^{2}}\left[(\gamma-1)p_\mathrm{(j)}v^\mathrm{(j)}v^\mathrm{(i)}-\gamma{p^\mathrm{(t)}}v^\mathrm{(i)}\right],
\end{eqnarray}
with the Lorentz factor $\gamma$ as measured by ZAMOs, the Carter momenta $p^{(\mu)}$ and the modulus of the velocity $v$.
Only the component $\hat{p^\mathrm{t}}$ is needed and one gets
\begin{eqnarray}
\hat{p^\mathrm{t}} & = & \gamma\left[p^\mathrm{(t)}-v^\mathrm{(r)}p_\mathrm{(r)}-v^\mathrm{(\theta)}p_\mathrm{(\theta)}-v^\mathrm{(\Phi)}p_\mathrm{(\Phi)}\right] \nonumber\\
& = &
\gamma\left[\frac{1}{\alpha}(1-\omega\lambda)-v^\mathrm{(r)}\frac{\sqrt{R}}{\rho\sqrt{\Delta}}-v^\mathrm{(\theta)}\frac{\sqrt{\Theta}}{\rho}-v^\mathrm{(\Phi)}\frac{\lambda}{\tilde{\omega}}\right].
\end{eqnarray}
With $\hat{p}^\mathrm{t}_\mathrm{obs}\rightarrow{1}$ for $r\rightarrow\infty$ as valid for a distant observer, the generalized Doppler factor in the emitter
frame is evaluated to
\begin{eqnarray} \label{eq:g}
g & = & \frac{\alpha_\mathrm{em}}{\gamma\left[(1-\omega\lambda)-\alpha{v^\mathrm{(r)}}\frac{\sqrt{\mathcal{R}_\mathrm{0}}}{\rho\sqrt{\Delta}}-\alpha{v^\mathrm{(\theta)}}\frac{\sqrt{\Theta}}{\rho}-\alpha{v^\mathrm{(\Phi)}}\frac{\lambda}{\tilde{\omega}}\right]_\mathrm{em}} \nonumber\\
& = & \frac{\alpha_\mathrm{em}}{\gamma\left[1-\alpha{v^\mathrm{(r)}}\frac{\sqrt{\mathcal{R}_\mathrm{0}}}{\rho\sqrt{\Delta}}-\alpha{v^\mathrm{(\theta)}}\frac{\sqrt{\Theta}}{\rho}-\lambda\Omega\right]_\mathrm{em}}.
\end{eqnarray}
\begin{figure}
  \rotatebox{0}{\includegraphics[height=6.9cm,width=9cm]{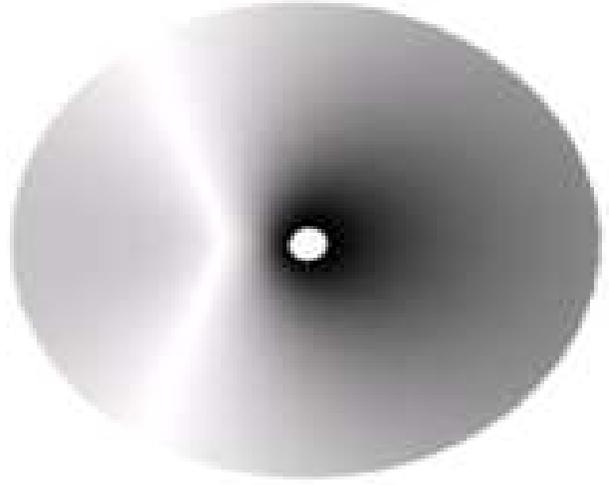}}
  \caption{Simulated disk image at inclination of $30^{\circ}$ and extreme Kerr $a=0.999999$ with color--coded distribution of generalized Doppler
  factor $g$ (white corresponds to $g=1$). The beamed blue approaching left side is clearly seen (triangular segment, $g>1$). Gravitational redshift (dark colors, $g<1$)
  darkens the inner edge until $g$ vanishes at the horizon.} \label{fig:gdistr}
\end{figure}
Here, the already introduced Boyer--Lindquist functions are used again. Furthermore, there is the plasma velocity field in the ZAMO--frame
(denoted by round brackets) $v^\mathrm{(r)}$,$v^\mathrm{(\theta)}$ and $v^\mathrm{(\Phi)}$, two polynomials
of the fourth order, $\mathcal{R}_\mathrm{0}$ and $\Theta$, associated with the Carter Constant and the specific angular momentum $\lambda = J/E$,
the ratio of the two conserved quantities along each photon path, angular momentum $J$ and total energy $E$. The polynomial $\mathcal{R}$ (in most
general form, see Eq. (\ref{R})) becomes for photons ($m=0$)
\begin{equation}
\frac{\mathcal{R}_\mathrm{0}}{E^{2}}=r^{4}+(a^{2}-\lambda^{2}-\mathcal{C})r^{2}+2\left[\mathcal{C}+(\lambda-a)^{2}\right]r-a^{2}\mathcal{C}.
\end{equation}
Usually, one assumes Keplerian rotation for $r\geq r_\mathrm{ms}$ and sets
\begin{equation} \label{vPhi}
v^\mathrm{(\Phi)}=\tilde{\omega}\left(\frac{\Omega-\omega}{\alpha}\right).
\end{equation}
The angular frequency, $\Omega$, has for radii greater than marginal stability a Keplerian profile
\begin{equation} \label{eq:kep}
\Omega = \Omega_\mathrm{K} = \pm\frac{\sqrt{M}}{\sqrt{r^{3}}\pm a\sqrt{M}}\
\ \mathrm{for} \ \ r > r_\mathrm{ms},
\end{equation}
(with upper sign for prograde, lower sign for retrograde orbits)
and usually one forgets about the contributions of $v^\mathrm{(r)}$ and $v^\mathrm{(\theta)}$. These components can nevertheless be important for the
emission line profile \citep{andy}. A more realistic and complicated velocity field than only Keplerian rotation will be discussed in the next section.
For $r \leq r_\mathrm{ms}$, {\em constant specific angular momentum} is assumed. This is motivated by 1D hydrodynamic accretion disk models \citep{jose} and
relativistic 2D hydrodynamic accretion disk models \citep{spindel}.
Here, one can set
\begin{equation}
\Omega = \Omega_\mathrm{in} =
\omega+\frac{\alpha^{2}}{\tilde{\omega}^{2}}\frac{\lambda_\mathrm{ms}}{1-\omega \lambda_\mathrm{ms}}\
\ \mathrm{for} \ \ r \leq r_\mathrm{ms},
\end{equation}
with specific angular momentum at the marginally stable orbit given by
\begin{equation}
\lambda_\mathrm{ms}=\frac{\tilde{\omega}^{2}_\mathrm{ms}\left(\Omega_\mathrm{K,ms}-\omega_\mathrm{ms}\right)}{\alpha^{2}_\mathrm{ms}+\omega_\mathrm{ms}\tilde{\omega}^{2}_\mathrm{ms}\left(\Omega_\mathrm{K,ms}-\omega_\mathrm{ms}\right),}
\end{equation}
(The index ms refers to the quantity to be taken at the radius of marginal stability, $r_\mathrm{ms}$.). \\
In Fig. \ref{fig:gdistr}, the simulated distribution of the Doppler factor $g$ from Eq. (\ref{eq:g}) on a disk with parameters typical
for Seyfert--1 galaxies (i = $30^{\circ}$) is shown. First, one discovers the usual separation in blueshifted and redshifted segments due to rotation of the disk.
The left--to--right symmetry is compared to the
\begin{figure}
  \rotatebox{0}{\includegraphics[height=6cm,width=9cm]{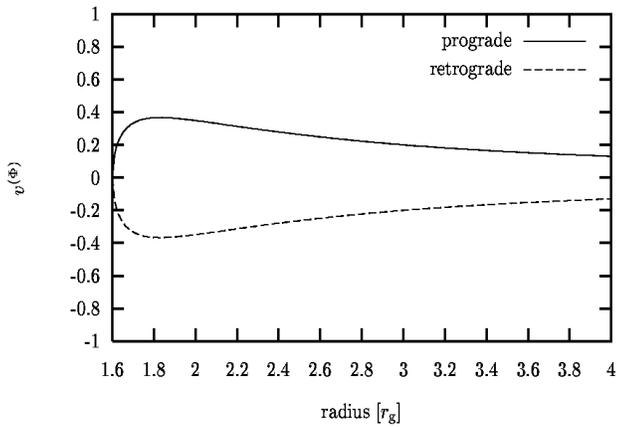}}
  \caption{Radial dependence of the velocity component $v^\mathrm{(\Phi)}$ in the equatorial plane for Kerr parameter $a=0.8$. Prograde case is shown in
  the upper curve; retrograde case is illustrated in the lower curve. The matter starts in free--fall from a drift radius of $R_\mathrm{t}=3.0 \ r_\mathrm{g}$.
  $v^\mathrm{(\Phi)}$ vanishes self--consistently (for ZAMOs) at the horizon at $r_\mathrm{H}=1.6 \ r_\mathrm{g}$.} \label{fig:v_phi}
\end{figure}
Newtonian case distorted. The left side of the disk rotates towards the observer and this radiation is beamed due to special relativistic effects:
the matter rotation velocity becomes comparable to the speed of light. This fact can be derived from Fig. \ref{fig:v_phi}, illustrating the prograde and
retrograde velocity component, $v^\mathrm{(\Phi)}$, in the ZAMO frame. The Keplerian profile of $v^\mathrm{(\Phi)}$ holds only until reaching the radius of marginal
stability, $r_\mathrm{ms}$. For smaller radii, one has to model this velocity component. The simplest way is to assume free--falling matter. In Fig. \ref{fig:v_phi}
the drift radius starts already at $R_\mathrm{t}=3.0 \ r_\mathrm{g}$. Then the test particle falls on geodesics onto the black hole. The velocity field is a complicated
superposition of $v^\mathrm{(\Phi)}$ and $v^\mathrm{(r)}$. Besides, the plot depicts that the component $v^\mathrm{(\Phi)}$ as observed by the ZAMO
reaches a magnitude that is comparable to the speed of light: symmetrically in both cases, prograde and retrograde rotation, the maximum rotation velocity
is about 0.4c! Additionally, in both cases, at the horizon, this velocity component drops down to zero.\\
For completeness, the distribution of $g$ down to the event horizon of the black hole is calculated to illustrate mainly the effect of gravitational
redshift. Certainly, a disk that touches the horizon is unphysical and would not be stable, unless in the case of maximum Kerr, $a=1.0$, where all
characteristic black hole radii coincide to $1 \ r_\mathrm{g}$. Near the horizon the $g$--factor (Eq. (\ref{eq:g})) significantly drops
down and vanishes at the horizon itself (or equivalently, the redshift $z$ becomes infinity): the curvature of the black hole grasps everything
that is near, even light.
\begin{figure}
  \rotatebox{0}{\includegraphics[height=7.63cm,width=9cm]{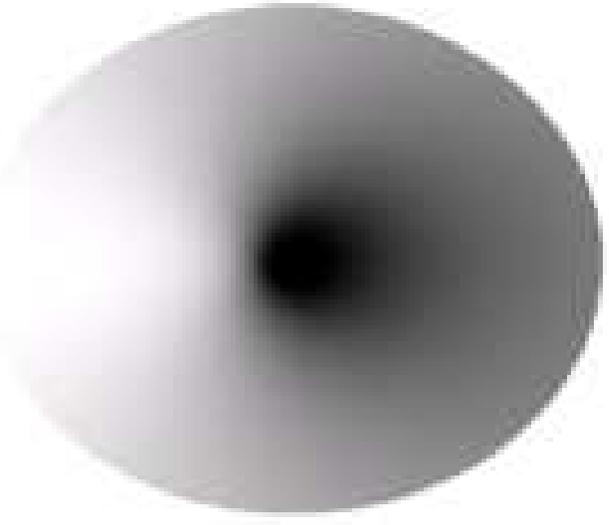}}
  \caption{Distribution of the Doppler factor $g^{4}$ over the disk. The inner disk edge touches at $r_\mathrm{in}=1.0015 \ r_\mathrm{g}$
  the horizon of the black hole with $a=0.999999$. The inclination angle is $30^{\circ}$ and the outer edge is at $r_\mathrm{out}=30.0 \ r_\mathrm{g}$.
  The {\em shadow} of the black hole is clearly seen. Beaming on the left approaching side brightens any emission originating there. The right
  receeding part of the disk is darkened. This distribution is always folded in the flux integral and therefore important for {\em any emission}
  near black holes.} \label{fig:g4distr}
\end{figure}
Fig. \ref{fig:g4distr} now shows the Doppler factor to the fourth power. In our approach, this is an essential ingredient {\em for any emission}
originating from the vicinity of a black hole (ray tracing using transfer functions has a weight to the third power after all). This weight to the
fourth power shows the relevance of the generalized Doppler factor. The $g$--factor significantly decreases inwards and becomes zero at the event horizon.
The $g$--factor to the fourth power shows therefore a dramatic tininess around the black hole. This feature
is called the \textbf{central shadow} of the black hole \citep{fal}. Likewise, one remarks a very bright region on the approaching part of the
disk (beaming). For midly and highly inclined disks there is a chance to measure this clear brightness step, e.g. with sub--mm VLBI in the Galactic Center.
This may be a method to prove the existence of the supermassive black hole with $3.7 \times 10^{6} M_{\odot}$ \citep{schoe} associated with the source
Sgr A$^{*}$ {\em by its radio emission}. In near future, \textbf{direct imaging} of black holes may be possible!
%
%
%
\section{The velocity field of the emitters} \label{sec:emit}
\begin{figure}
  \rotatebox{0}{\includegraphics[height=5.82cm,width=9cm]{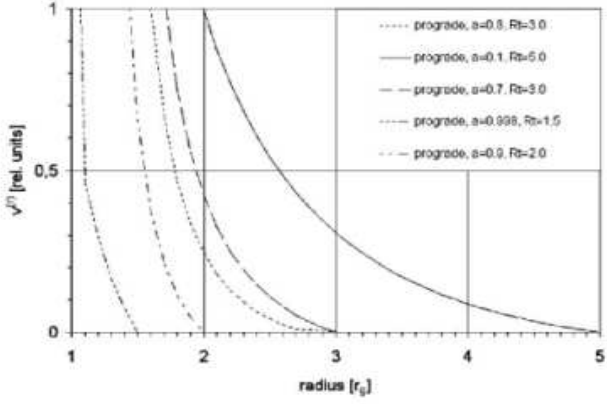}}
  \caption{Radial velocity component in the ZAMO for different radial drift models, that means different rotation state of black hole, $a$, and
  drift radius, $R_\mathrm{t}$. In {\em all} cases, the radial drift becomes light speed at the horizon radii.} \label{fig:v_rvar}
\end{figure}

\begin{figure}
  \rotatebox{0}{\includegraphics[height=14cm,width=9cm]{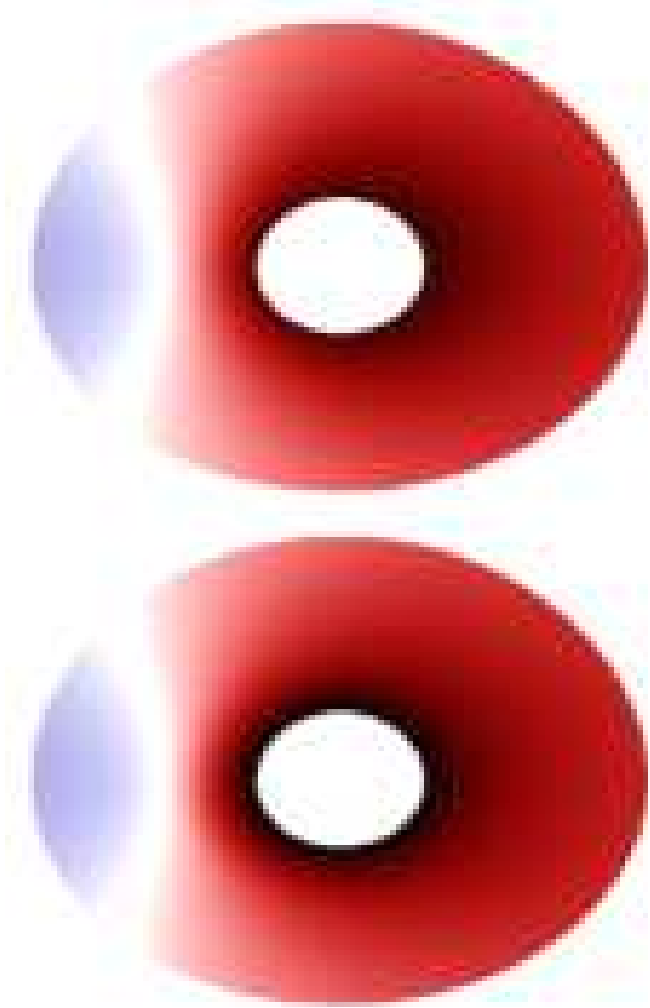}}
  \caption{Influence of radial drift. Top: pure Keplerian rotation $v^\mathrm{(\Phi)}\not=0, \ v^\mathrm{(r)}=0$.
  Bottom: Keplerian rotation \textbf{plus} radial drift $v^\mathrm{(\Phi)}\not=0, \ v^\mathrm{(r)}\not=0$.
  \textbf{Enhanced gravitional redshift} is clearly seen by darkening near the horizon.
  Parameters were $a=0.1$, $i=40^{\circ}$, $r_\mathrm{in}=r_\mathrm{H}=1.996 \ r_\mathrm{g}$, $r_\mathrm{out}=10.0 \ r_\mathrm{g}$.
  The radial drift starts at $R_\mathrm{t}=5.0 \ r_\mathrm{g}$} \label{fig:radvskep}
\end{figure}

The generalized Doppler factor $g$ depends on the plasma motions (here written in the ZAMO frame),
$g=g(v^\mathrm{(r)},v^\mathrm{(\Theta)},v^\mathrm{(\Phi)})$. The usual issue forgets
about the radial and poloidal components and derives line profiles
from emitting plasma with only Keplerian motion, i.e. pure rotation. Recent non--radiative 3D--MHD 
pseudo--Newtonian simulations \citep{BH} again confirm this inflow and, additionally, the poloidal motion 
of disk winds. Therefore a more complicated plasma kinematics near the event horizon of Kerr black holes
is investigated in this section. The behaviour of the velocity components $v^\mathrm{(r)}$, $v^\mathrm{(\Theta)}$
and $v^\mathrm{(\Phi)}$ at the horizon is presented in detail below. \\
The radial velocity component will be determined by the accretion process. In this paper, it is considered that the \textbf{radial drift} starts 
at a certain radius, the \textbf{truncation radius} $R_\mathrm{t}$, that can even be {\em greater} than the radius for marginal stability $r_\mathrm{ms}$.
Truncation is an essential feature of radiative accretion theory. This is motivated by radiative hydrodynamical accretion models of \citep{huj},
discussed in Sect. \ref{sec:xrayspec}. The correct free--fall behaviour in the Kerr geometry starting at radius $R_\mathrm{t}$ can be
studied by introducing the radial velocity component
\begin{equation} \label{vr}
v^\mathrm{r} = \frac{dr}{dt}=\frac{dr}{d\tau}\frac{d\tau}{dt} = \frac{\sqrt{\mathcal{R}}}{\rho^{2}}\frac{\alpha^{2}}{1-\omega\lambda} = \frac{\Delta}{\Sigma^{2}}\frac{\sqrt{\mathcal{R}}}{1-\omega\lambda},
\end{equation}
where $\mathcal{R}$ denotes the radial potential \citep{cart}. This polynomial generally depends on Carter's
constant $\mathcal{C}$, specific angular momentum $\lambda$, total energy $E$ and mass of the plasma
test particle $m$,
\begin{equation} \label{R}
\frac{\mathcal{R}}{E^{2}}=\left[(r^{2}+a^{2})-a\lambda\right]^{2}-\Delta\left[\frac{\mathcal{C}}{E^{2}}+(\lambda-a)^{2}+r^{2}\frac{m^{2}}{E^{2}}\right].
\end{equation}
Carter's constant $\mathcal{C}$ vanishes in the equatorial plane. \\
\begin{figure}
  \rotatebox{0}{\includegraphics[height=4.96cm,width=9cm]{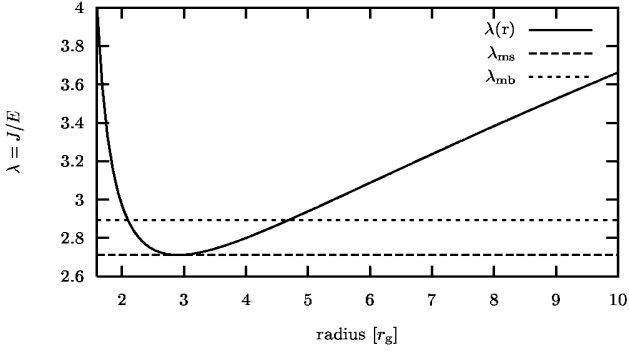}}
  \caption{Distribution of specific angular momentum, $\lambda\mathrm{(r)}$, for Kerr parameter
  $a=0.8$. Fixing a constant specific angular momentum, $\lambda_\mathrm{t}$, is only allowed
  inbetween the two levels: the lower one, $\lambda_\mathrm{ms}$, and the upper one, $\lambda_\mathrm{mb}$.} \label{fig:lamb}
\end{figure}
$v^\mathrm{r}$ is in a normalized ZAMO frame (indicated by round brackets)
\begin{equation}
v^\mathrm{(r)}=\frac{\exp(\mu_\mathrm{r})}{\alpha}=\frac{\Sigma}{\Delta}v^\mathrm{r}.
\end{equation}
This quantity has to be evaluated in the equatorial plane, $\theta
= \frac{\pi}{2}$. The specific angular momentum, $\lambda=J/E$, follows from
\begin{eqnarray} \label{EJ}
\frac{E}{m} & = & \frac{r^{2}-2Mr\pm a\sqrt{Mr}}{r\sqrt{r^{2}-3Mr\pm2a\sqrt{M r}}}, \\
\frac{J}{m} & = & \pm\sqrt{M r}\frac{r^{2}\mp2a\sqrt{M
r}+a^{2}}{r\sqrt{r^{2}-3Mr\pm2a\sqrt{M r}}},
\end{eqnarray}
(upper sign prograde, lower sign retrograde orbits). \\
This provides the radial drift
\begin{equation} \label{eq:vr}
v^\mathrm{(r)}=\frac{\sqrt{\mathcal{R}}}{\Sigma(1-\omega\lambda)}.
\end{equation}
The \textbf{kinematics at the event horizon},
\begin{equation} \label{eq:rH}
r_\mathrm{H}=M+\sqrt{M^{2}-a^{2}},
\end{equation}
can be explored by evaluating the Boyer--Lindquist functions
\begin{eqnarray}
& \Delta(r_\mathrm{H}) = 0, & \alpha(r_\mathrm{H}) = 0, \nonumber\\
& \Sigma_{\pi/2}(r_\mathrm{H}) = 2r_\mathrm{H}, & \tilde\omega_{\pi/2}(r_\mathrm{H}) = 2, \\
& \omega(r_\mathrm{H}) = a/2r_\mathrm{H} \equiv\Omega_\mathrm{H}. \nonumber
\end{eqnarray}
This finally yields
\begin{eqnarray} \label{eq:vr1}
v^\mathrm{(r)}(r_\mathrm{H}) & = &
\frac{1}{2r_\mathrm{H}(E-\frac{a}{2r_\mathrm{H}}J)}\left[E(1+2\sqrt{1-a^{2}}+1)-aJ\right]
\nonumber\\
& = & \frac{1}{2r_\mathrm{H}E-aJ}\left[2E(1+\sqrt{1-a^{2}})-aJ\right] \nonumber\\
& = & 1.
\end{eqnarray}
Therefore, matter in the equatorial plane enters the horizon with the
\textbf{speed of light}. This is a general feature of the Kerr geometry and
{\em fully independent} of any accretion model. \\
A realistic approach to model the radial drift is to fix the
conservatives $E$ and $J$ when reaching the truncation radius
$R_\mathrm{t}$ with the values that they take at this radius.
Then, the particle will fall freely into the black hole. The disk is
truncated at $R_\mathrm{t}\geq r_\mathrm{ms}$, the truncation radius
gives the conservatives, $E(R_\mathrm{t})$ and $J(R_\mathrm{t})$. Otherwise, the 
radius of marginal stability fixes the conservatives to $E(r_\mathrm{ms})$ and $J(r_\mathrm{ms})$. \\
\begin{figure}
  \rotatebox{0}{\includegraphics[height=5.26cm,width=9cm]{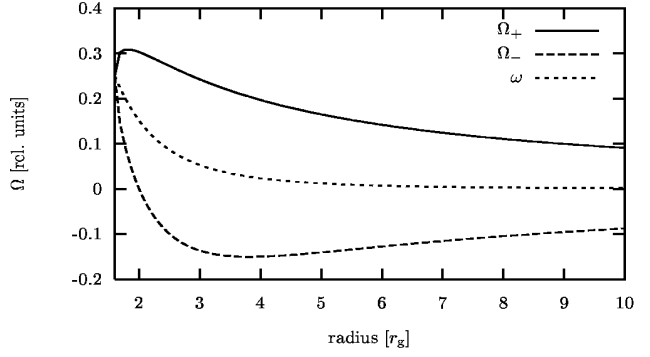}}
  \caption{Radial distribution of prograde (solid curve), $\Omega_{+}$, and retrograde (dashed curve)
  angular velocity limit, $\Omega_{-}$, for Kerr parameter $a=0.8$. The dotted curve illustrates
  the frame--dragging potential, $\omega$. At the horizon, here at $r_\mathrm{H}=1.6 \ r_\mathrm{g}$, all curves
  coincide. This is the visualization of the frame--dragging effect.} \label{fig:omegalimit}
\end{figure}
It is \textbf{not} possible to give arbitrary values for
constant $\lambda_\mathrm{t}$, that means to fix arbitrary $R_\mathrm{t}$. As has been
discussed in torus solutions on the Kerr geometry \citep{abram},
there is only an allowed interval of $\lambda_\mathrm{ms}\leq \lambda_\mathrm{t}\leq
\lambda_\mathrm{mb}$ for the specific angular momentum, where ms denotes the
orbit of marginal stability and mb the marginally bound orbit.
This allowed interval $\lambda_\mathrm{t}\in[\lambda_\mathrm{ms},\lambda_\mathrm{mb}]$ is depicted
in Fig. \ref{fig:lamb} for a Kerr parameter of $a=0.8$. The general distribution
of the specific angular momentum in the Kerr geometry is represented by the curve.
The drift radius has to be chosen so that its according specific angular momentum, $\lambda_\mathrm{t}$,
just lays inbetween the two horizontal lines. These lines belong to the specific angular momenta
at the radius of marginal stability, $\lambda_\mathrm{ms}$, and at the marginally bound radius,
$\lambda_\mathrm{mb}$. The physical reason for this restriction is, that $\Omega$
satisfies the condition
\begin{equation}
\Omega_{-}\leq\Omega\leq\Omega_{+} \ , \
\Omega_{\pm}=\omega\pm\frac{\alpha}{\tilde\omega}.
\end{equation}
These two limits are shown in Fig. \ref{fig:omegalimit} for $a=0.8$. It must be considered in modelling the
radial drift with constant specific angular momentum that the resulting $\Omega$ lays inbetween these
two curves, $\Omega_{+}$ and $\Omega_{-}$. Otherwise, the model is invalid. This is because a
global observer is globally time--like,
$U_{\alpha}U^{\alpha}\leq0$. By the frame--dragging effect, this
observer becomes light--like, because the horizon is a light--like
surface ({\em null surface}). If one chooses arbitrary specific
angular momenta, one can run into trouble and injure the above
allowed stripe for $\Omega$. In general, a slow rotating Kerr
black hole, e.g. low Kerr parameter $a$, allows higher truncation
radii. This can be investigated in Fig. \ref{fig:v_rvar} where models with different
radial drift are compared. The models differ in Kerr parameter, $a$, and drift radius, $R_\mathrm{t}$, where the radial velocity 
component, $v^\mathrm{(r)}$, starts to become finite. As can be seen, the radial velocity increases very steeply
to smaller radii and becomes {\em always} the speed of light as has been proven analytically in Eq.
(\ref{eq:vr1}). This component is the additive that changes the $g$--factor in Eq. (\ref{eq:g})
significantly. The essential consequence is that the gravitational redshift effect is enforced
what can be directly investigated in Fig. \ref{fig:radvskep}: the inner edge shows a clear darkening due
to smaller $g$--factors. \\
Let us now consider the toroidal velocity component, $v^\mathrm{(\Phi)}$.
From the Carter momenta it follows that
\begin{equation} \label{vphi}
v^\mathrm{(\Phi)}=\frac{\tilde\omega}{\alpha}(\Omega-\omega).
\end{equation}
Another general feature of the Kerr metric is the {\em frame--dragging
effect}. Anything must co--rotate with the black holes
horizon when reaching $r_\mathrm{H}$
\begin{equation} \label{OmegaH}
\Omega(r_\mathrm{H})\equiv\Omega_\mathrm{H}=\omega.
\end{equation}
This can be immediately deduced from the general formula valid in the Kerr metric
\begin{equation}
\Omega = \omega+\frac{\alpha^{2}}{\tilde{\omega}^{2}}\frac{\lambda}{1-\omega\lambda},
\end{equation}
and the fact that $\alpha(r_\mathrm{H})=0$. So, the azimuthal velocity in the ZAMO frame
{\em vanishes in general at the horizon} as can be seen from Eq. (\ref{vphi}).
Fig. \ref{fig:v_phi} confirms this for $a=0.8$ and drift radius $R_\mathrm{t}=3.0$ \\
The poloidal velocity component, $v^\mathrm{\Theta}$, in the ZAMO frame
\begin{equation}
v^\mathrm{\Theta}=\frac{d\theta}{dt}=\frac{d\theta}{d\tau}\frac{d\tau}{dt}=\frac{\sqrt{\Theta}}{\rho^{2}}\frac{\alpha^{2}}{1-\omega\lambda}
\end{equation}
vanishes in the equatorial plane, hence the polynomial $\Theta$ defined by
\begin{equation}
\frac{\Theta}{E^{2}}=\frac{\mathcal{C}}{E^{2}}-\left[a^{2}\left(\frac{m^{2}}{E^{2}}-1\right)+\lambda^{2}\mathrm{cosec}^{2}\theta\right]\cos^{2}\theta
\end{equation}
vanishes for $\theta=\pi/2$. Therefore, the poloidal velocity component can be neglected for cold, infinitely
thin standard disks. It may have relevance for slim or thick disks, when
calculating their line flux. However this requires volume ray tracing, covariant radiation
transfer and more physics as for standard disks what lies beyond the scope of this paper.
%
%
%
%
\section{Emissivity models} \label{sec:emiss}
\begin{figure}
  \rotatebox{0}{\includegraphics[height=6.98cm,width=9cm]{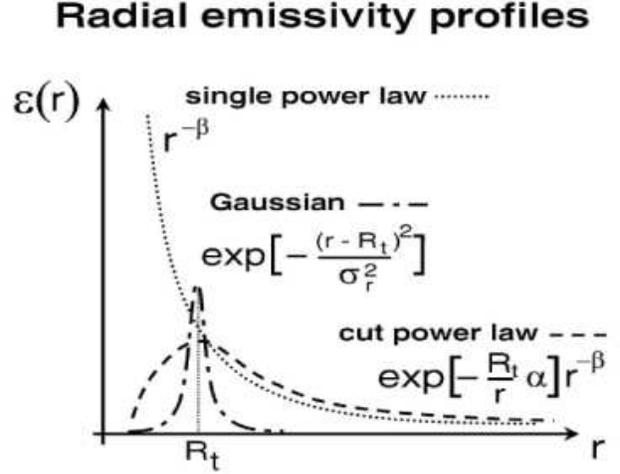}}
  \caption{Illustration of different radial emissivity models} \label{fig:emmod}
\end{figure}
In general, the disk emissivity is the quantity that has to be folded into the flux integral to evaluate the observed flux from accretion
disks. One applies the emissivity as kind of {\em profile function} that has a radial and potentially an angular dependence. The
resulting spectra (continuum flux, line flux) have therefore no physical flux units but arbitrary units. \\
The classical approach to model emissivities of accretion disks arose in the relativistic generalization of (non--relativistic) SSDs \citep{PT,NT}.
Here, the radial emissivity profile follows a \textbf{single power law} with slope index $\beta = 3.0$
\begin{equation}
\epsilon(r)\propto r^{-\beta}.
\end{equation}
This was motivated by the physics of standard accretion disks where other physical quantities follow this
power law behaviour, too. Classical standard disks truncate at the radius of marginal stability, $r_\mathrm{ms}$,
where consequently the emission of the disk breaks off naturally. In the historical issue \citep{PT} the {\em Zero--torque boundary
condition} (ZTBC) has been applied. Then, the emissivity decreases steeply at this characteristic radius. Modern accretion theory resigns
the ZTBC \citep{MF}. On the one hand, this was theoretically motivated by efficient mechanisms to produce MHD turbulence, the well--known
magneto--rotational instability (MRI, \citet{BH2}). The MRI and the coupling of magnetized accretion disk to the black hole magnetosphere induces
magnetically driven torques that injure the ZTBC. On the other hand, observations suggest significantly deviating disk emissivity profiles,
$\epsilon(r)\propto r^{-4.5}$, giving a hint to higher dissipation at radii $r\leq r_\mathrm{ms}$ \citep{wilms}. \\
The classical ansatz, $\beta=3$, was generalized later to a \textbf{single power law} index different from 3.0. The index $\beta$ may vary locally on
the standard disk because different illumination from the hot corona and complex structure of the ionization layer on the disk can cause changes
in the emissivity: $\beta$ becomes a fit parameter in X--ray spectroscopy. Another aspect is, that at $r\gtrsim r_\mathrm{ms}$ the disk may be truncated.
The inner accretion disk topology depends mainly on the total mass accretion rate as investigated in hydrodynamical accretion theory.
For high accretion rate, the disk extends nearly to the black hole horizon and the ADAF \citep{naryi} is small. For low accretion rate, the ADAF is
larger: a spheroidal evaporated hot region is attached to the standard disk and forms a sandwich configuration. A variable accretion rate is
responsable for the different spectral states (very high, high, intermediate, low, quiescent) and radially oscillating transition radius between
SSD and ADAF \citep{esin,jose}.\\
In the TSD scenario that emerges only in radiative accretion theory, it is possible that
the matter then decouples from the standard disk in free--falling clouds. This plasma packages drift radially inwards and are advected from the black
hole. This is major motivation of our new plasma velocity field model presented in Sect. \ref{sec:emit}. This phenomenon justifies an emissivity
profile that does {\em not} decrease as sharp at radii comparable to $r_\mathrm{ms}$. \\
Furthermore, one can test so--called \textbf{double power laws} or \textbf{broken power laws} where at a certain radius the power law index jumps
to another value. \\
In a simple approach, it is easy to modify the classical emissivity profile in the first step by a \textbf{cut--power law}.
The most simple model would envisage a direct vertical cut in the emissivity profile. This may be over-simplified because it is expected that the
emissivity is enhanced at the inner truncation due to stronger illumination by the corona. An even more adequate model considers the
single power law with a modification by an \textbf{exponential factor} that suppresses the emission at smaller radii. Then, the emissivity profile
decreases more smoothly at the truncation radius. {\em Physically}, this is motivated by pseudo--Newtonian, radiative two--component hydrodynamics
accretion disk models in 2.5D axisymmetry \citep{huj} where Comptonization, synchrotron radiation, bremsstrahlung and ion/electron conduction are
considered. These simulations have shown that \textbf{disk truncation} at radii $r\gtrsim r_\mathrm{ms}$ seems to hold. It is supposed that this
scenario is realized in AGN, miqroquasars and GBHCs. \\
Then, the radial emissivity in cut--power law form satisfies
\begin{equation} \label{eq:cut}
\epsilon(r)\propto \exp\left[-\frac{R_\mathrm{t}}{r}\alpha\right] \ r^{-\beta},
\end{equation}
with {\em truncation radius} $R_\mathrm{t}$, steepness parameter $\alpha$ and single power law index $\beta$.
The maximum of this distribution can be found at $r_\mathrm{max}=R_\mathrm{t}\alpha/\beta$. If one assumes a standard emissivity power law, $\beta=3.0$, for
the second factor, one can fix $\alpha=3.0$, too. Then it is ensured that the emissivity peaks at $R_\mathrm{t}$ and exponentially decreases at smaller
radii. This disk emissivity profile is comparable to the historical work \citep{PT} but exhibits a higher emissivity at the inner disk edge. This is
plausible because X--ray irradiation may be enhanced due to better illumination for geometrical reasons.\\
But there are alternatives to model the emissivity profile. In a new approach, a {\em localized emissivity} realized
by a \textbf{Gaussian emissivity profile} is assumed
\begin{equation} \label{eq:gauss}
\epsilon(r)\propto \exp\left[-\frac{(r-R_\mathrm{t})^{2}}{\sigma_\mathrm{r}^{2}}\right].
\end{equation}
One can easily control the width of the Gaussian by the parameter $\sigma_\mathrm{r}$. Narrow and broad Gaussians with constant $\sigma_\mathrm{r}$ are
investigated. Another possibility is variable width by assuming $\sigma_\mathrm{r} = \eta R_\mathrm{t}$. Here the width scales with the truncation radius
$R_\mathrm{t}$, $\eta$ denotes a constant factor that can be suitably chosen. \\
Gaussian emissivity profiles serve for modelling emission from a ring. In principle, the ring is the {\em Greens function} of the axisymmetric emission
problem in the Kerr geometry. One can simply imagine that any spectral shapes originating from extended axisymmetric and flat emitters are
composed of small rings, e.g. infinitely narrow Gaussians. Hence, one can produce a relativistic emission line profile by superimposing the flux
data of several rings. Therefore, the Gaussian emissivity profile is well--suited to study different emitting regions.\\
Fig. \ref{fig:emmod} illustrates all radial emissivities presented here in direct comparison.
\begin{figure}
  \rotatebox{0}{\includegraphics[height=9.2cm,width=9cm]{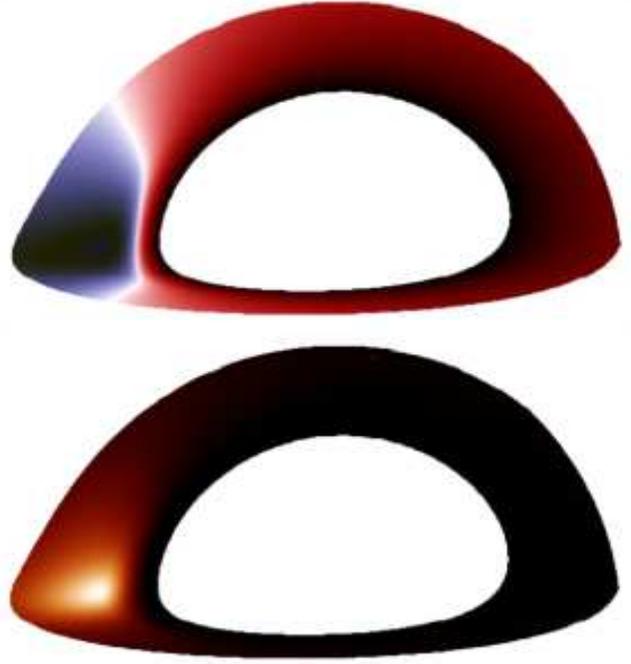}}
  \caption{The upper disk shows the distribution of the $g$--factor over a narrow ring with radial drift starting at $R_\mathrm{t}=3.0 \ r_\mathrm{g}$.
  The inclination angle is $80^{\circ}$, the black hole rotation is fast, $a=0.8$, inner edge is the horizon at $r_\mathrm{in}=r_\mathrm{H}=1.6 \ r_\mathrm{g}$
  and outer edge is only at $r_\mathrm{out}=4.6 \ r_\mathrm{g}$. The lower disk indicates the associated distribution of the emission of the same narrow ring.
  Here, the emissivity follows a broad Gaussian profile with $R_\mathrm{t}=3.0 \ r_\mathrm{g}$ and constant Gaussian width $\sigma_\mathrm{r}=3.0$.}
  \label{fig:gemdisk_i80}
\end{figure}
%
%
%
%
\section{Visualization of standard disks} \label{sec:visdisk}
With these preparations, the investigation of the emission distributed over the disk is straight--forward. One very interesting case is to study
the influence of the radial drift $v^\mathrm{(r)}\not=0$, as presented in Sect. \ref{sec:emit}, Eq. (\ref{eq:vr}).
This new ansatz to model the velocity field results mainly in an enhancement of the gravitational redshift process (compare Fig.
\ref{fig:radvskep}).\\
Another instructive example are \textbf{high--inclined disks}. Surely, this can not be applied to AGN type--1 and AGN type--2 are obscured by the large--scale
dust torus, so that there is few chance to observe broad X--ray emission lines from type--2 AGN. But maybe this can be applied to stellar black holes in
suitable orientation that can be found in microquasars and GBHCs.
Besides, these studies are interesting to understand GR effects in this part of the parameter space.\\
A narrow ring laying directly around a Kerr black hole horizon is examined. Fig. \ref{fig:gemdisk_i80} now shows firstly the distribution of the generalized Doppler
factor $g$ from Eq. (\ref{eq:g}) under consideration of a Keplerian velocity component, $v^\mathrm{(\Phi)}$, and a drift component, $v^\mathrm{(r)}$ that is
switched on at truncation radius $R_\mathrm{t}=3.0 \ r_\mathrm{g}$. Generally, for highly inclined disks the beaming of the approaching part is very strong. This illustrates
the deep blue spot at the left disk segment in Fig. \ref{fig:gemdisk_i80}.
The left approaching part of the disk exhibits a clear brightness step as can be investigated in the emission distributed over the disk. Apparently,
this is a consequence of the $g$--factor jumping from high values in the beaming feature to very small values at gravitational redshift feature respective
the  horizon. If observers may find a possibility to look at high--inclined standard disks of cosmic black hole candidates, this step would be a
distinct observable feature confirming the Kerr geometry\footnote{More images of disk distributions and emission line profiles can be found at
http://www.lsw.uni-heidelberg.de/users/amueller}.
%
%
%
%
\section{Relativistic emission lines: calculation and basic features} \label{sec:feline}
Let us start with the calculation of the line flux from the rendered disk images. Traditionally in such calculations, it has been assumed that
the emitting volumes are embedded into the surface of the disk and determine the generalized Doppler factor for each pixel in the image. So,
in principle, one has to integrate over the disk images (that show the emission on each pixel), because a distant observer is not able to
resolve the emitting disk. Physically, the observed spectral flux is evaluated generally by
\begin{equation} \label{eq:obsfl}
F^\mathrm{obs}=\int_\mathrm{image}d\Xi \ I_{\nu}^\mathrm{obs},
\end{equation}
with the observed intensity $I_{\nu}^\mathrm{obs}$ and an element of solid angle $d\Xi$, covering a pixel of the disk as seen by a distant observer.
The observed intensity into the rest frame of the emitting plasma (denoted by $\hat{}$ ) holds
\begin{equation}
I_{\nu}^\mathrm{obs}=g^{3}\hat{I}_{\nu}^\mathrm{em},
\end{equation}
with the generalized Doppler factor, $g$, as introduced in Eq. (\ref{eq:g}).
In the emitter frame, one has to make an ansatz for the intrinsic line shape. Usually, a Dirac delta function suffices to describe the transition at
rest frame frequency $\nu_{0}$. One can assume a Gaussian profile motivated by Compton broadening, too. This is only one part, the frequency dependence.
The other part is the \textbf{radial disk emissivity} that satisfies one of the models elaborated in Sect. \ref{sec:emiss}. In our work
it is consecutively assumed that
\begin{equation} \label{emflux}
\hat{F}_{\nu}^\mathrm{em}=\pi \hat{I}_{\nu}^\mathrm{em} = \epsilon(r)\delta\left(\nu_\mathrm{em}-\nu_{0}\right).
\end{equation}
Finally, one has to consider a special feature of the delta distribution
\begin{equation}
\delta(\nu_\mathrm{em}-\nu_{0})=\delta((\nu_\mathrm{obs}-g\nu_{0})/g)=g\delta(\nu_\mathrm{obs}-g\nu_{0}).
\end{equation}
So, plug all this into the flux integral (\ref{eq:obsfl}) one gets:
\begin{equation} \label{flux}
F_\mathrm{obs}(E_\mathrm{obs})=\int_\mathrm{image}\epsilon(r)g^{4}\delta(E_\mathrm{obs}-gE_{0})d\Xi,
\end{equation}
with observed flux, $F_\mathrm{obs}$, line emission energy in the local rest frame, $E_{0}$, observed line energy in the observers frame, $E_\mathrm{obs}$,
and solid angle element, $d\Xi$. Eq. (\ref{flux}) shows that the flux is not a trivial integral over the disk but a complicated convolution of intrinsically
emitted line weighted with the generalized Doppler factor to the fourth power and the radial emissivity function. \\
Line profiles are simulated that depend in general on the parameter set $\{{a,i,r_\mathrm{in},r_\mathrm{out},v^\mathrm{(\Phi)},v^\mathrm{(r)},v^\mathrm{(\Theta)},\epsilon(r)}\}$,
with the Kerr parameter $a$, the inclination angle of the disk $i$, the inner $r_\mathrm{in}$ and outer edge $r_\mathrm{out}$ of the thin standard disk, the plasma
velocity field in the ZAMO frame $v^\mathrm{(\Phi)},v^\mathrm{(r)},v^\mathrm{(\Theta)}$ as elaborated in Sect. \ref{sec:emit} and the emissivity law $\epsilon(r)$ that can
depend on one, two or three parameters itself, as has been shown in Sect. \ref{sec:emiss}.\\
Therefore, one deals with a rather huge parameter space that results in a variety of emission line shapes. Consideration of astro--chemistry makes it
even more difficult, because different species and transitions (Fe K$\alpha$, Fe K$\beta$, Ni K$\alpha$, Cr K$\alpha$ etc.) can contribute to
the observed X--ray line feature at approximately 6.5 keV. \\
Generally, there are \textbf{three effects} \citep{fab3} that influence the emission line profile\footnote{Thermal broadening has a tiny contribution and can be
neglected.}: the \textbf{Doppler effect}, already known from classical Newtonian emitting disks, forms a {\em characteristic double horn structure}
at medium to high inclinations. At low inclinations, the profile is triangular. The Doppler feature is for relativistic accretion disks distorted. \\
\textbf{Beaming} is a special relativistic effect. The rotating matter moves relativistically fast and the radiation is collimated (beamed) in direction of
motion. Beaming intensifies the blue wing of emission lines with respect to the red wing. This effect is extraordinary important for disks with inner
edges close to the event horizon because there the toroidal velocity as seen by ZAMOs is comparable to the speed of light. At the horizon itself, the
toroidal velocity steeply decreases to become zero at $r_\mathrm{H}$ as shown in Fig. \ref{fig:v_phi}. \\
\textbf{Gravitational redshift} is a general relativistic effect. Matter and radiation near the black hole feels the strong
curvature of space--time and looses energy when escaping this highly--curved region. The consequence is a shift of all photons that succeed in escaping the
black holes sphere of action to the red branch of the spectrum. This effect results in an elongated red tail of the line. Besides, it lowers the
relic red Doppler peak with respect to the blue wing. \\
The {\em frame--dragging effect} in the Kerr space--time forces anything to co--rotate. In particular, at the rotating horizon anything co--rotates
with the angular velocity $\Omega_\mathrm{H}=\omega$ dictated by the black hole, compare Eq. (\ref{OmegaH}).
This means, that photons that do {\em not} rotate in the black holes direction may be forced to turn back. In Schwarzschild space--times, $a=0$,
rotating matter is forced to {\em stop rotation} at the horizon, building up a boundary layer. \\
\begin{figure}
  \includegraphics[height=6cm,width=6cm]{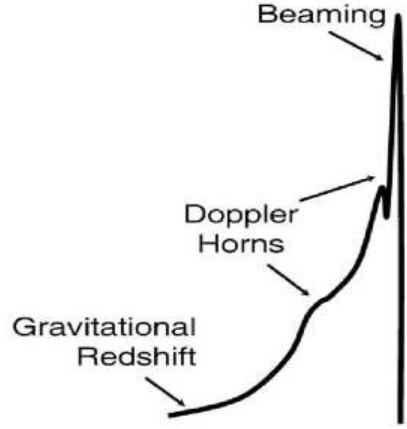}
  \caption{Three effects form the line profile. This simulated profile of an iron K line has typical parameters for Seyfert--1s. The observer is intermediately inclined, the plasma rotates
  only Keplerian and typical disk sizes $r_\mathrm{in} = r_\mathrm{ms}$, $r_\mathrm{out} = 50 \ r_\mathrm{g}$ were taken.} \label{fig:line}
\end{figure}
The combination of all these effects deform the spectral emission line to a very skewed profile with typically a long red tail (gravitational redshift),
an intense beamed blue peak (beaming) and a double--peaked structure (Doppler boosting) as illustrated in Fig. \ref{fig:line}. But the
topology of the line depends {\em strongly} on the inclination angle and the emissivity law (see Sect. \ref{sec:emiss}). \\
The parameters used to fit an relativistic emission line profile are numerous. In general, the strongest influence on both, broadness and topology, has
the inclination of the disk. At low inclinations, the observer does not feel anything of the rotation of the disk and stares onto a nearly face--on
oriented disk. There is no significant velocity component towards the observer. The emission line profile is {\em triangular}. This changes dramatically
at high inclinations: the observer sees the approaching and receeding part of the rotating accretion disk. One can easily fix the inclination in a first
approximation by determining the high--energy cut--off at the blue wing of the observed line. The emission line profile is {\em double--horned}. \\
Another aspect is the rotation of the black hole. For physical reasons, it seems clear that the Schwarzschild solution is ruled out. From stellar
evolution theory it is known that stellar black holes formed by gravitational collapse of a rotating progenitor stars. Angular momentum can not radiated away
by emission of gravitational waves. Therefore, at least stellar black holes are supposed to rotate. Even though the formation process of supermassive
black holes is still enigmatic, one can assume that they form by merging of stellar Kerr black holes ({\em hierarchical growth}) or other rotating
constituents. Then, it is very likely that black holes in general carry angular momentum. And even if black holes do not rotate from the beginning
of their formation, the accretion mechanism makes sure that the Kerr solution is the approbiate metric: matter carries angular momentum and winds up
the black hole spin when accreted (compare Sect.\ref{sec:raytr}). \\
But the precise rotation state is unsure: it is not clear whether to take a Kerr parameter of 0.5 or 0.998. It may be a good approach to assume high
rotation near extreme Kerr ($a \approx 0.99$) because these 'elder black holes' accrete already for a long time and are wind--up strongly. Besides, the
relativistic iron lines suggest that the emission originates from the innermost regions of Seyfert galaxies. A Kerr parameter of $a = 0.5$ is associated
with an orbit of marginal stability of 4.23 $r_\mathrm{g}$. This may be to far away to cause the measured gravitational redshift
effects. Disk truncation softens this argument as elaborated in Sect. \ref{sec:xrayspec}. \\
As can be seen in the equation of the generalized Doppler factor (\ref{eq:g}), the velocity field of the radiation emitting matter is very important for
the resulting emission line profiles. Typically, Keplerian motion (see Eq. (\ref{eq:kep})) has been taken into account \citep{fab}. Here, only the
component $v^\mathrm{(\Phi)}$ is non--vanishing whereas $v^\mathrm{(\Theta)}$ and $v^\mathrm{(r)}$ are zero. But it is certainly essential to
incorporate a \textbf{radial drift}, e.g. a free--fall motion that is comparable to the modulus of the Keplerian
motion when reaching the orbit of marginal stability. Here the correct fully relativistic free--fall formula in the Kerr geometry, Eq.
(\ref{eq:vr}), is adapted. \\
Poloidal components may also contribute. There is strong evidence that the jets of AGN and the "blobs" of GBHCs are {\em magnetically driven} and form
in the direct neighborhood of the black holes horizon. The interaction of the (ergospheric) accretion disk with the black hole magnetosphere drives
torsional Alfv$\acute\mathrm{e}$n waves and finally an outflow (wind) that is collimated on larger scales to form jets. Certainly, the jets of Seyferts are weak
and this argument holds rather for another type of AGN, the Quasars. Besides, to adequately consider
poloidal motion, the ray tracer has to be modified to a {\em volume ray tracer} that renders the emission of 3D objects. But this has to be coupled to
\textbf{radiative MHD on the Kerr geometry}, which is still an unresolved problem. \\
Recently, the {\em variable} iron line profiles from pseudo--Newtonian MHD accretion theory have been calculated \citep{AR}. This model mimics the
Schwarzschild geometry and shows nicely the dependence of the line flux from accretion but does still not incorporate relativistic effects of a Kerr
black hole. This issue hints into the right direction of emission line simulations: coupling of accretion theory and X--ray spectroscopy. Another progress
has been done to simulate non--radiative accretion flows on the Kerr geometry \citep{DeVH}. The next step will be to connect Relativity and radiation
transfer to study radiatively cooled accretion flows on the background of the Kerr geometry. Here, in our first investigation, poloidal velocity components
of the plasma are neglected. Firstly, the influence of radial drift and alternative emissivity models are studied. \\
One further important aspect is the disk size, especially the inner edge. The outer edge is rather irrelevant for classical emissivities, $\beta\geq 2.0$
\citep{RN}. The inner and outer radius of the disk are fixed and afterwards the image of this disk is determined. The
historical approach is based on the famous SSD model and considers a thin and cold standard disk that extends outwards to several hundreds
gravitational radii ($\sim 10^{2} \ r_\mathrm{g} $) and inwards to the marginally stable orbit, $r_\mathrm{ms}$. For a rapidly rotating black hole ($a = 0.998$),
$r_\mathrm{ms} = 1.23 \ r_\mathrm{g}$ is very close to the horizon $r_\mathrm{H} = 1.063 \ r_\mathrm{g}$! Then gravitational redshift influences strongly the radiation from the
inner disk edge. The accretion disk is supposed to be {\em not} as close to the black hole. This is motivated by pseudo--Newtonian radiative hydrodynamics
simulations that show for typical accretion rates of Seyfert galaxies, $\dot M = 0.01 \ M_{\odot}/\mathrm{yr}$, that the disk is \textbf{truncated}
at $r\geq r_\mathrm{ms}$. Typical truncation radii $R_\mathrm{t}$ are 10 to 15 $r_\mathrm{g}$, depending on the Kerr parameter $a$ and the allowed stripe introduced in Fig.
\ref{fig:lamb} and \ref{fig:omegalimit}.
%
%
%
%
\section{Line zoo: topological classification and parameter space} \label{sec:linzoo}
\begin{figure}
\begin{center}
  \rotatebox{-90}{\includegraphics[height=7cm,width=4.57cm]{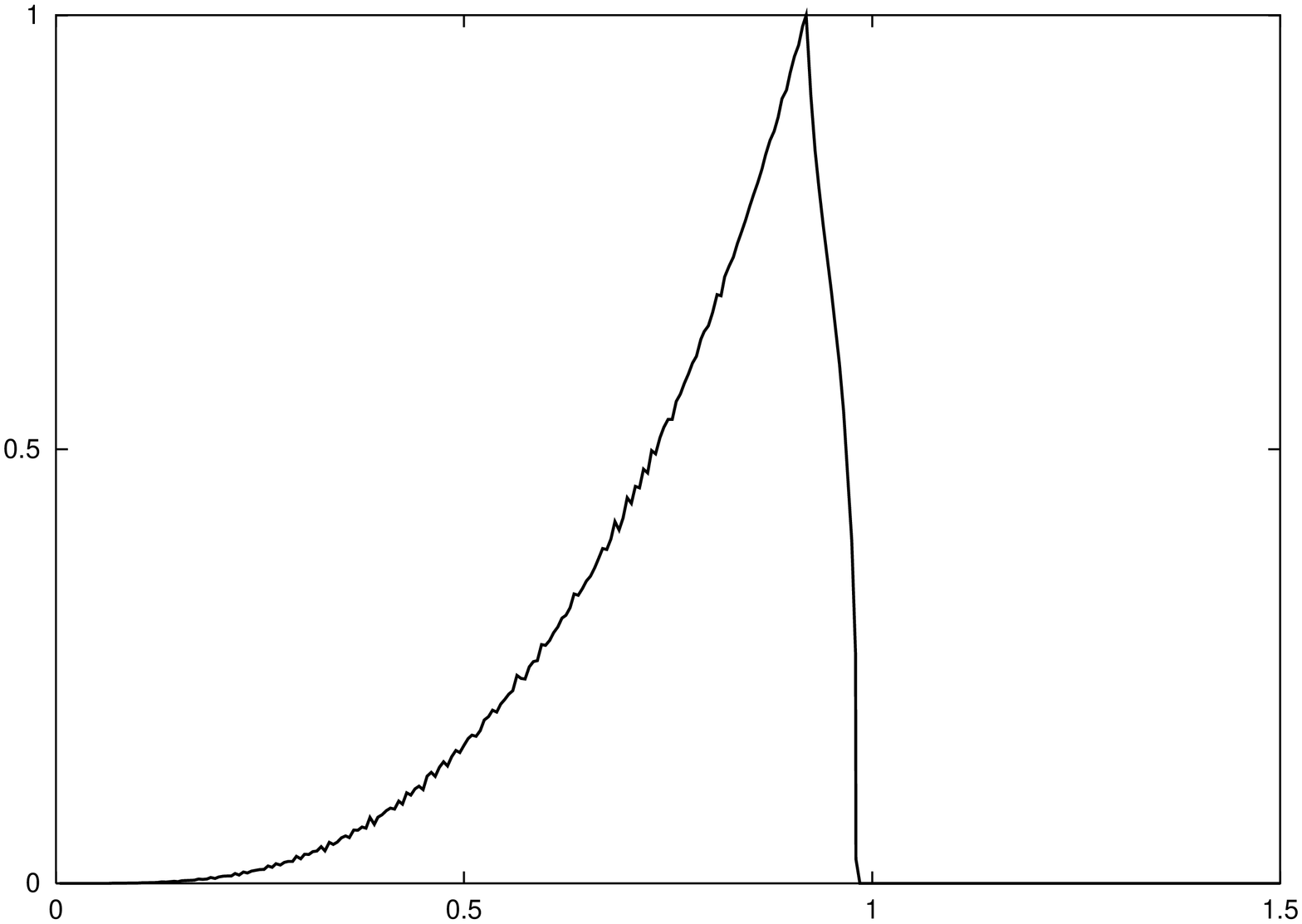}}
  \rotatebox{-90}{\includegraphics[height=7cm,width=4.57cm]{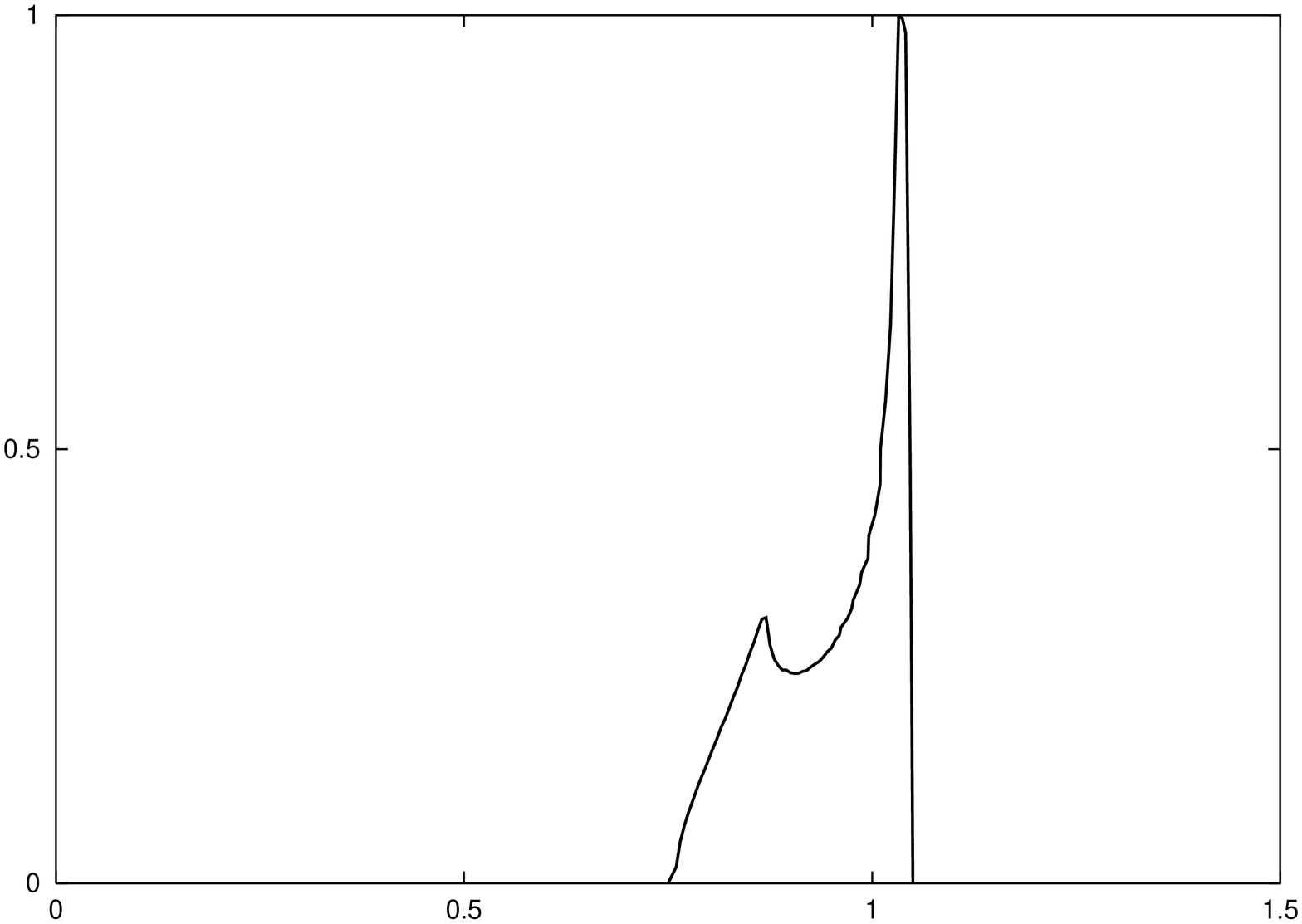}}
  \rotatebox{-90}{\includegraphics[height=7cm,width=4.57cm]{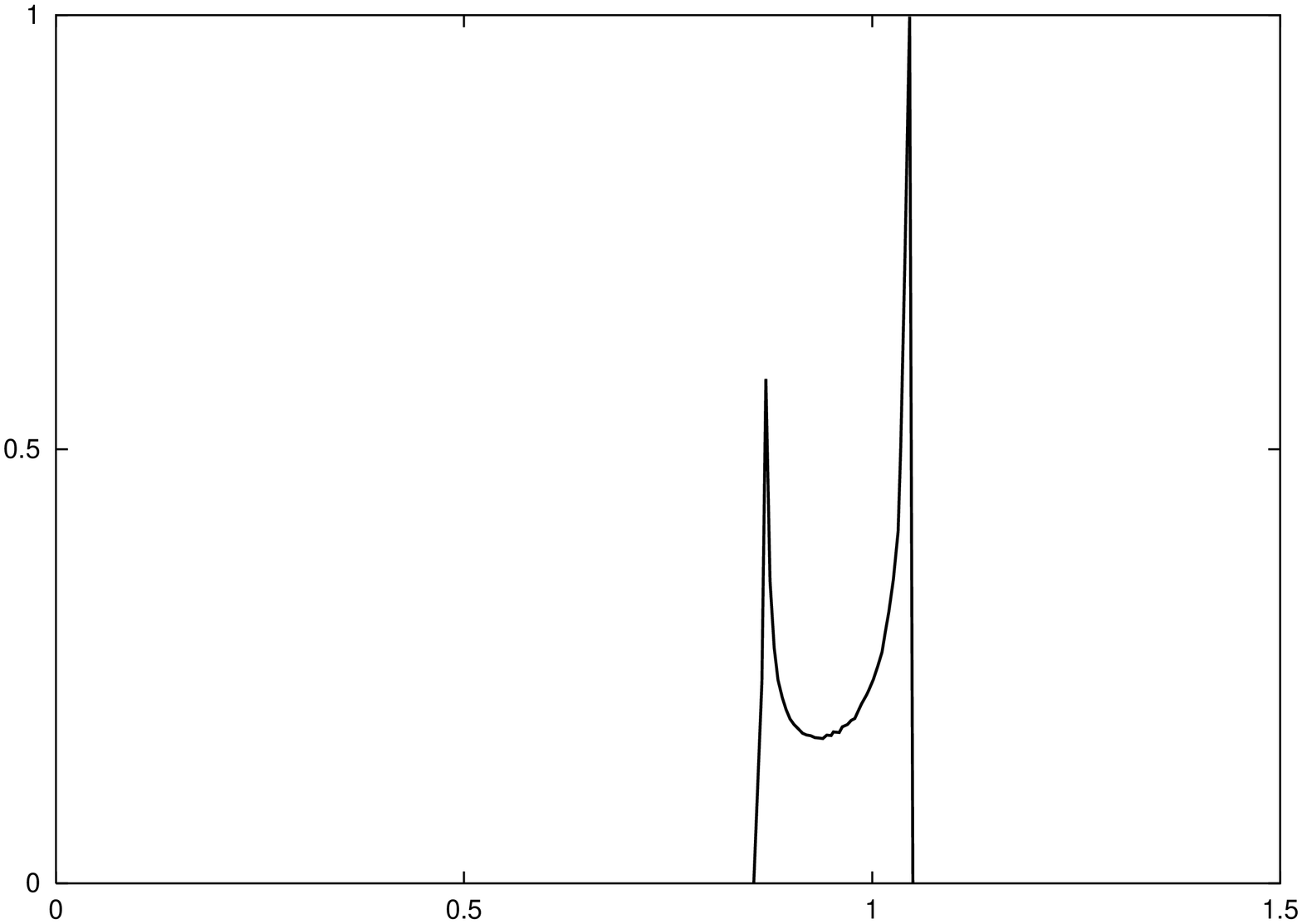}}
  \rotatebox{-90}{\includegraphics[height=7cm,width=4.57cm]{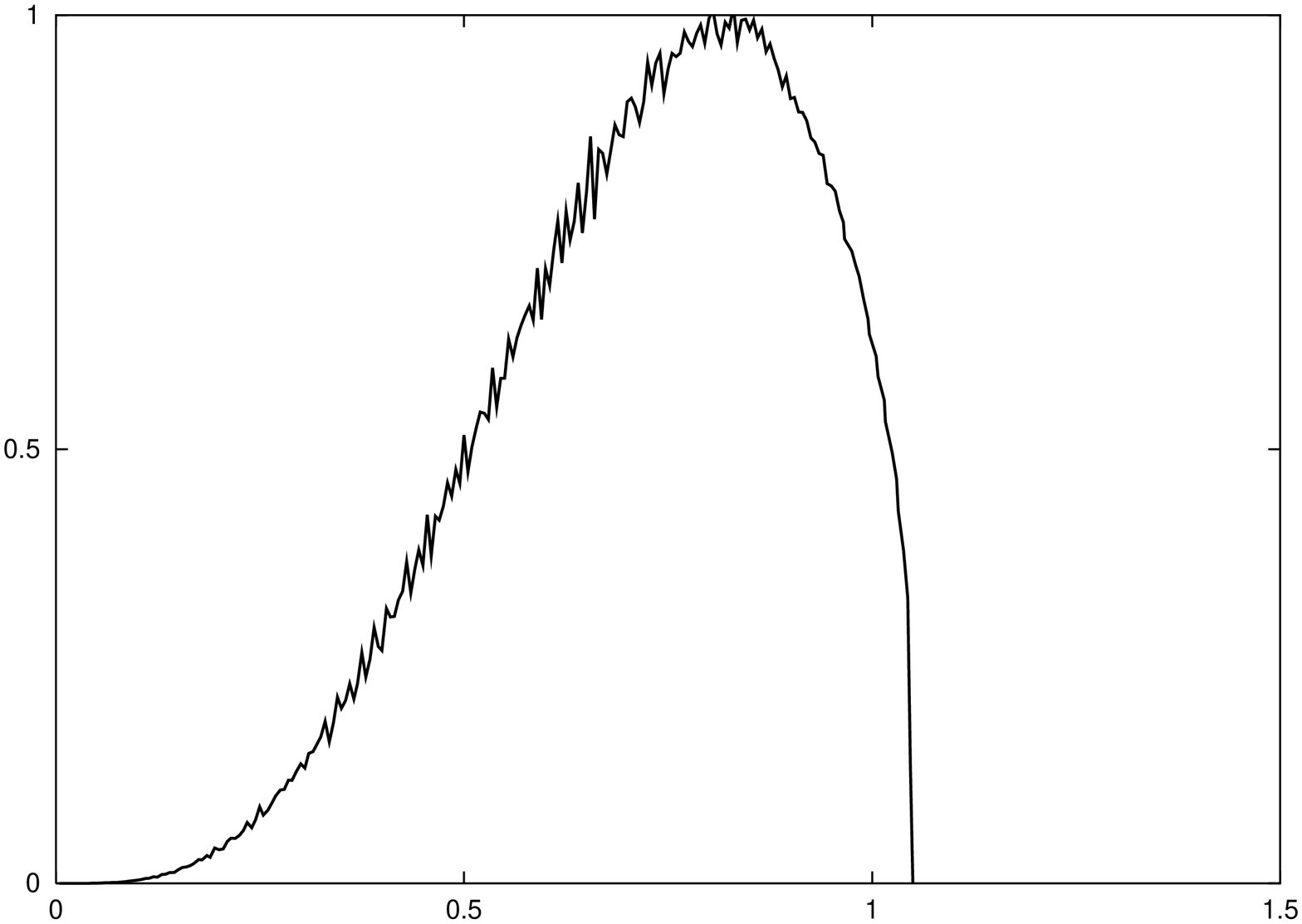}}
  \rotatebox{-90}{\includegraphics[height=7cm,width=4.57cm]{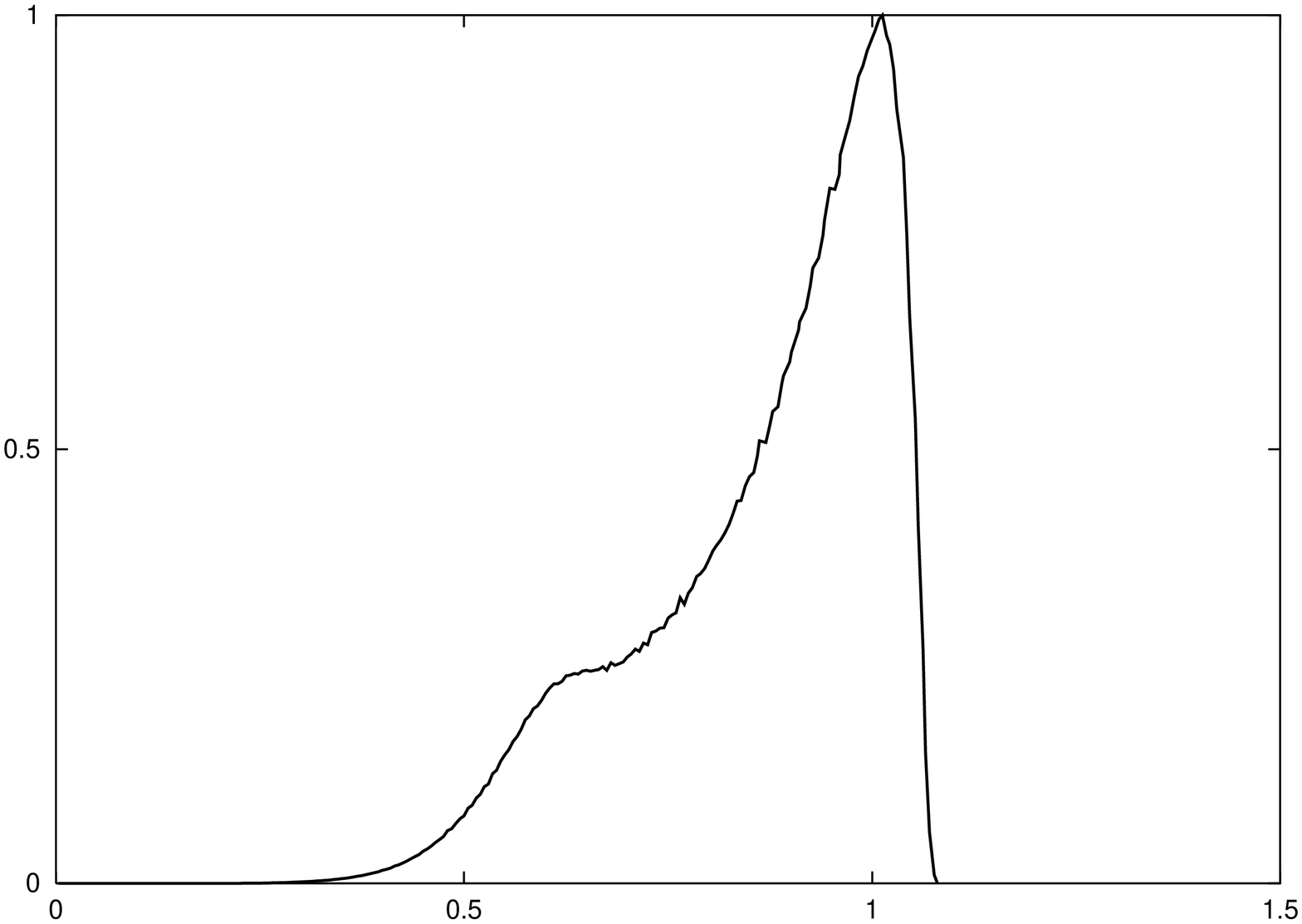}}
\end{center}
  \caption{A selection of topology types of relativistic emission lines. From top to bottom: triangular, double--horned,
  double--peaked, bumpy and shoulder--like. Line flux in normalized arbitrary units is plotted over $g$-factor.} \label{fig:zoo}
\end{figure}

\begin{figure}
  \rotatebox{0}{\includegraphics[height=8.27cm,width=9cm]{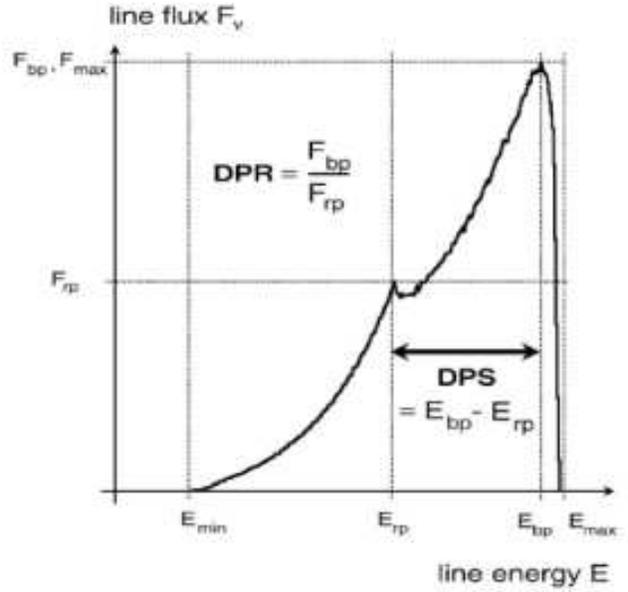}}
  \caption{Illustration of line criteria to determine characteristics of a typical emission line profile. The position of the relic Doppler peaks
  (if existing) can be used to fix their flux ratio (Doppler Peak Ratio, $\mathrm{DPR}$) and energetic distance (Doppler Peak Spacing, $\mathrm{DPS}$).
  Other specific quantities are the maximal flux, $F_\mathrm{max}$, and the minimum and maximum energy of a line , $E_\mathrm{min}$ and $E_\mathrm{max}$. These criteria
  can be used to describe a line shape, observed or theoretically derived, with a few numerical values.} \label{fig:lincrit}
\end{figure}

\begin{figure}
  \rotatebox{-90}{\includegraphics[height=9cm,width=5.79cm]{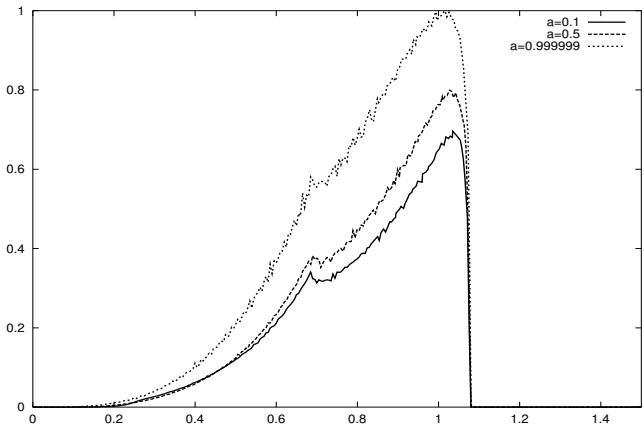}}
  \caption{Investigation of the dependence on the Kerr parameter $a$. The plasma motion is only Keplerian, the inclination is $40^{\circ}$. The black
  hole spin decreases from top to bottom line profile, $a=0.999999, \ 0.5, \ 0.1$. The inner disk edge is equal to the corresponding black hole
  horizon, $r_\mathrm{in}=r_\mathrm{H}(a)$, and the disk edges are chosen so that the total disk surface remains constant.} \label{fig:astudy}
\end{figure}

\begin{figure}
  \rotatebox{0}{\includegraphics[height=4.78cm,width=9cm]{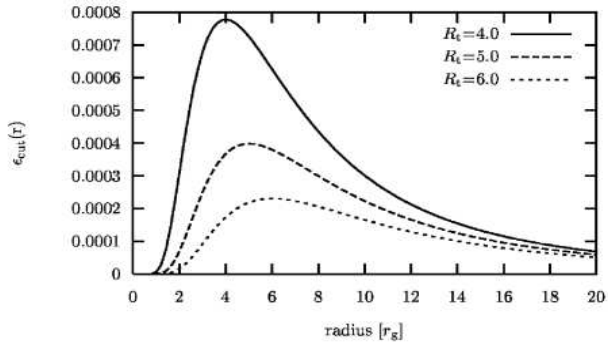}}
  \caption{Cut--emissivity with $\alpha=\beta=3.0$ for different truncation radii. From bottom to top, the drift radius increases,
  $R_\mathrm{t}=4.0, \ 5.0, \ 6.0$. With this choice of $\alpha$ and $\beta$, the emissivity peaks naturally at $R_\mathrm{t}$. For $r\leq R_\mathrm{t}$
  the emissivity decreases exponentially.} \label{fig:cutem456}
\end{figure}

\begin{figure}
  \rotatebox{-90}{\includegraphics[height=9cm,width=5.17cm]{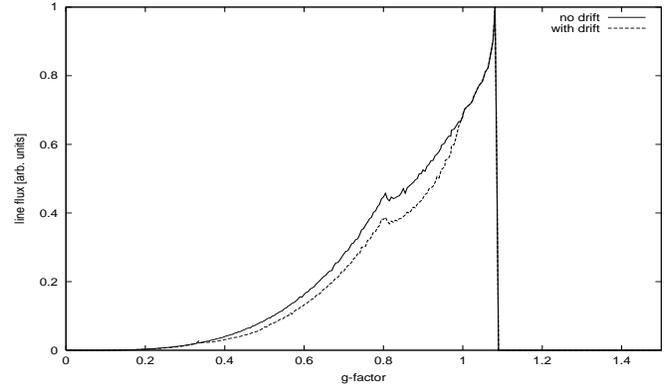}}
  \caption{Direct confrontation of relativistic emission lines with and without radial drift model. A single--power law emissivity with $\beta=3.0$
  is chosen. The black hole spin parameter is $a=0.8$, the inclination angle is $40^{\circ}$, $r_\mathrm{in}=r_\mathrm{H}=1.6 \ r_\mathrm{g}$ and
  $r_\mathrm{out}=20.0 \ r_\mathrm{g}$. The pure Keplerian case, top line, is modelled by constant specific angular momentum for
  $r\leq r_\mathrm{ms}=2.91 \ r_\mathrm{g}$. The non-Keplerian case, bottom line, includes drift starting at $R_\mathrm{t}=4.5 \ r_\mathrm{g}$.} \label{fig:Kepvsdrift}
\end{figure}

\begin{figure}
  \rotatebox{0}{\includegraphics[height=5.17cm,width=9cm]{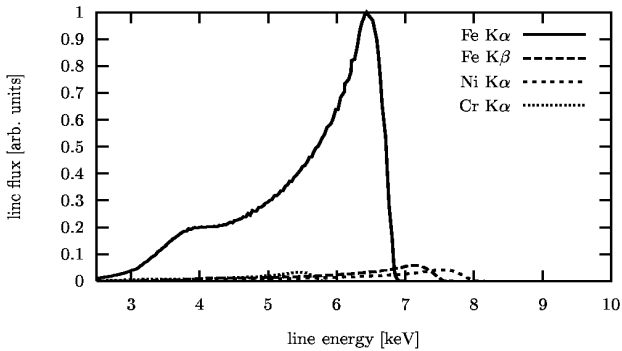}}
  \caption{Suggestion of a multi--species line system fitting the XMM--Newton observation of MCG--6--30--15 (Fabian et al., 2002). A line
  complex consisting of Fe K$\alpha$ at 6.4 keV, Fe K$\beta$ at 7.06 keV, Ni K$\alpha$ at 7.48 keV and one weak contribution of Cr K$\alpha$
  at 5.41 keV (all energies measured in the rest frame) is simulated.} \label{fig:Fabfit}
\end{figure}

In this section, {\em all} types of line topologies emerging in our simulations are presented. From the plain line form, one can
derive the following \textbf{nomenclature of relativistic emission lines}
\begin{itemize}
\item triangular,
\item double--horned,
\item double--peaked,
\item bumpy,
\item shoulder--like.
\end{itemize}
Fig. \ref{fig:zoo} illustrates all prototypes of this terminology. As can be seen from the parameters, the \textbf{triangular}
form follows from low inclination angles at rather high emissivity. This is trivial, because the Doppler
effect is suppressed for nearly face--on observed disks: the red peak of the line vanishes. \\
\textbf{Double--horned} line shapes are somewhat standard profiles, because many astrophysical objects exhibit these typical form. Everything needed
for that is a Keplerian velocity field, intermediate inclination, $i\approx 30^{\circ}$, and a standard single power law to reproduce a line profile with
two Doppler boosted horns, where the blue one is beamed as usual.\\
The \textbf{double--peaked} profiles are well--known from the Newtonian case of higher--inclined radiating disks. Here, the width of the relic Doppler
peaks is lower as compared to double--horned profiles. Mainly, this characteristic shape is a consequence of the space--time that is sufficiently
flat, so that the typical red tail from gravitational redshift lacks. This can be theoretically reproduced by shifting the inner edge of the disk 
outwards. The simulations with stepwise shifting show nicely the "motion" of the red tail until it decreases as sharp as the blue edge. A relatively
flat space--time is already reached around $25.0 \ r_\mathrm{g}$ corresponding to 0.25 AU for a typical Seyfert galaxy with $10^{6} \ \mathrm{M}_{\odot}$. 
Weakly accreting Seyfert galaxies with distant inner disk edges may fit this scenario. \\
\textbf{Bumpy} profiles are mainly caused by steep emissivity profiles. These emissivities cut away the emission from outer disk regions, especially
the beaming segment on the disk as depicted in Fig. \ref{fig:gdistr}. Therefore, the line profile lacks the characteristic sharp blue beaming
peak. The recent observation of Seyfert galaxy MCG--6--30--15 in the low state \citep{wilms} serves as an example of bumpy shapes. \\
\textbf{Shoulder--like} profiles exhibit a typically curved red wing. This feature is very sensitive to the parameters chosen. Our simulations showed,
that only relatively narrow Gaussian, broken or cut--emissivities could produce such a distinct feature. The motivation to produce red shoulders arose
from another observation of the Seyfert--1 galaxy
MCG--6--30--15 \citep{fab2}, now in the high state, showing such an extraordinary line shape. Blue shoulders can be produced by significant outflows, e.g.
non--vanishing poloidal plasma velocity components \citep{andy}. Small shoulder-like features attached to the blue wing are typically interpreted as other species
different from Fe K$\alpha$ such as Fe K$\beta$ \citep{fab2} or Ni K$\alpha$ \citep{wang}. These transitions have rest frame energies above 7 keV. \\
Frequently, X--ray astronomers observe narrow but apparently separated lines that are superimposed to broad line profiles. The interpretation often given is
that the two components originate from largely separated region: the broad line forms in the vicinity of the black hole whereas the narrow one is a
reflection of X--rays (coming from the central engine) at the large dust torus located on the kpc scale in AGN. Indeed, the dust harbors the relevant
species Fe, Ni, etc. Therefore, these features should {\em not} be observed at microquasars (and stellar black holes in general) in default of a
dust torus configuration. In simulations, it is possible to produce these \textbf{narrow peak composites} by adding an additional narrow Gaussian line
profile {\em without} relativistic broadening. \\
After this topological classification, one can start to analyze the line characteristics numerically. Fig. \ref{fig:lincrit} illustrates a few
\textbf{line criteria} that can be proposed to fix the line shape features. The utilisation of these criteria may simplify the comparison of observed and
theoretically derived emission line profiles. Taking the absissa as a guideline, the minimum energy of the line, $E_\mathrm{min}$, the energy of
the red relic Doppler peak (if visible), $E_\mathrm{rp}$, the energy of the blue relic Doppler peak (if visible), $E_\mathrm{bp}$, and finally the maximum
energy of the line, $E_\mathrm{max}$, are introduced. The two Doppler peaks may serve to determine their energetic distance, the \textbf{Doppler
Peak Spacing}:
\begin{equation} \label{DPSE}
\mathrm{DPS}=E_\mathrm{bp}-E_\mathrm{rp},
\end{equation}
or more generally in units of the generalized Doppler factor
\begin{equation} \label{DPSg}
\mathrm{DPS}=g_\mathrm{bp}-g_\mathrm{rp}.
\end{equation}
On the spectral flux axis, $F_\mathrm{rp}$ fixes the relic red and $F_\mathrm{bp}$ the relic blue Doppler peak.
To infer an unit--independent quantity, one can calculate the flux ratio of these two and get the \textbf{Doppler Peak Ratio}:
\begin{equation} \label{DPR}
\mathrm{DPR}=F_\mathrm{bp}/F_\mathrm{rp}.
\end{equation}
\begin{figure}
  \rotatebox{-90}{\includegraphics[height=9cm,width=5.87cm]{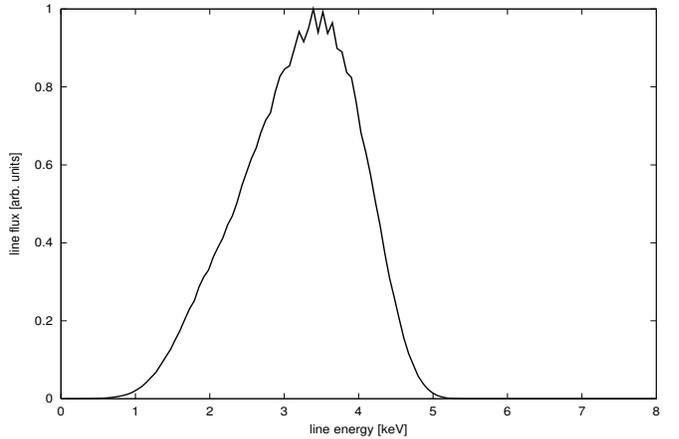}}
  \caption{Relativistic emission line forming directly in front of the horizon. The parameters were chosen to $i=30^{\circ}$, Kerr parameter $a=0.998$,
  disk size $r_\mathrm{in}=r_\mathrm{H}=1.0633 \ r_\mathrm{g}$, $r_\mathrm{out}=30.0 \ r_\mathrm{g}$, radial drift starting at
  $R_\mathrm{t}=1.5 \ r_\mathrm{g}$ and a localized emissivity by means of a Gaussian
  profile with $R_\mathrm{t}=1.5 \ r_\mathrm{g}$ and $\sigma_\mathrm{r}=0.4$. This more or less unphysical example for line emission should demonstrate
  the strongly suppressed emission due to the shadow of the black hole. Even if emission lines could form in the direct vicinity of black holes, the
  high power of the generalized Doppler factor $g$ reduces the flux and shifts the line significantly to lower energies due to gravitational redshift
  effects.} \label{fig:suppemiss}
\end{figure}
Additionally, the line reaches a maximum flux, $F_\mathrm{max}$, that may often match $F_\mathrm{bp}$ due to the beaming effect. \\
Another essential criterion that is even accessible by observation is the \textbf{line width}, in principle the total area of the emission
line weighted by the energy where the line peaks, $E_{0}$. In many cases, $E_{0}$ may coincide with $E_\mathrm{bp}$ because beaming determines the maximum
flux at the blue peak. The line width is evaluated numerically
\begin{equation} \label{LW}
\mathrm{LW}=\frac{1}{E_{0}}\sum_{i}F_\mathrm{i}\ \Delta E,
\end{equation}
where $\Delta E$ is equal to the numerical spectral resolution of the emission line. More generally, one can express this in units of the
$g$--factor to be independent of the line energy,
\begin{equation} \label{LWg}
\mathrm{LW}=\frac{1}{g_{0}}\sum_{i}F_\mathrm{i}\ \Delta g,
\end{equation}
with the corresponding $g$--factor $g_{0}=E^\mathrm{obs}_{0}/E^\mathrm{em}_{0}$. In both cases, constant spectral resolution $\Delta E_\mathrm{i}=\Delta E_{i+1}=...=\Delta E$
respectively $\Delta g_\mathrm{i}=\Delta g_{i+1}=...=\Delta g$ is assumed. Please consider that this line width has to be distinguished from the
observationally accessible equivalent width.\\
Now, only a few selected parameter studies are presented because as mentioned before the parameter space is huge. The first phenomenon to explore is
the frame--dragging effect or equivalently the depence of the line shape on the \textbf{Kerr parameter $a$}. Fig. \ref{fig:astudy}
shows only three lines at constant inclination angle, $i=40^{\circ}$, and classical emissivity power law, $\beta=3.0$, with variable Kerr parameter $a=0.999999$ (maximum Kerr),
$a=0.5$ (intermediate case) and $a=0.1$ (close to Schwarzschild). The plasma kinematics is exclusively determined by Keplerian rotation. The inner disk
edge is always coupled to the according horizon radius (Eq. (\ref{eq:rH})) and the outer edge is chosen so that the net area of the emitting
disk is always constant. Then, the line profiles are compared and one can state that the flux grows with increasing $a$. This is because $r_\mathrm{in}=r_\mathrm{H}$
decreases with increasing $a$ and the emissivity $\epsilon(r)\propto r^{-3.0}$ hence increases slightly extending to lower radii. This feature is
therefore somewhat artificial, but may be justified by enhanced dissipation at the inner disk edge. In this case, the emissivity profile is supposed
to contribute even for radii $r\leq r_\mathrm{ms}$. One can also state a changing line topology in this study: the ratios of the relic Doppler peak fluxes, $\mathrm{DPR}$,
as introduced in Eq. (\ref{DPR}), $\mathrm{DPR}(a=0.1)\approx 2.0$ and $\mathrm{DPR}(a=0.999999)\approx 1.7$, suggest that beaming is enhanced when the the black hole
rotates faster. This is plausible by considering the increasing toroidal velocity, $v^\mathrm{(\Phi)}$ respectively $\Omega$, with decreasing radius.\\
In Fig. \ref{fig:suppemiss} demonstrates \textbf{line dampening} by the {\em shadow of the black hole}. Here the radial drift model is applied, too, but
a pure Keplerian velocity field does not change the following argument. The Gaussian emissivity shape provides a suitable tool to easily fix a small
emitting ring laying around the horizon. The applied Gaussian emissivity peaks at $R_\mathrm{t}=1.5 \ r_\mathrm{g}$ and
vanishes in principle for radii $r>2.0 \ r_\mathrm{g}$. Therefore, the line form is nearly symmetric. This examples depicts that even if line formation
could happen within some $r_\mathrm{g}\lesssim r_\mathrm{ms}$, the line feature would be strongly suppressed by gravitational redshift. As analytically
proven in Sect. \ref{sec:raytr}, the $g$--factor vanishes at the horizon. Therefore, the powers of $g$ (third or fourth, depending on ray tracing technique)
folded into the flux integral reduce the observed flux. The consequence is a high--redshifted and damped emission line. As can be deduced from
Fig. \ref{fig:suppemiss}, a Fe K$\alpha$ line would peak at an observed energy of approximately only 3.4 keV. The line is significantly shifted to lower
energies due to gravitational redshift. Obviously, it would be hard to observe this weak line. But in any case, the fluorescence process is supposed to be
not possible in this proximity to the black hole because the accretion flow is too hot. The line dampening due to gravitational redshift is a strong
competitive effect against models with steep emissivity single power laws as proposed in \citet{wilms}.\\
\begin{figure}
  \rotatebox{-90}{\includegraphics[height=9cm,width=5.79cm]{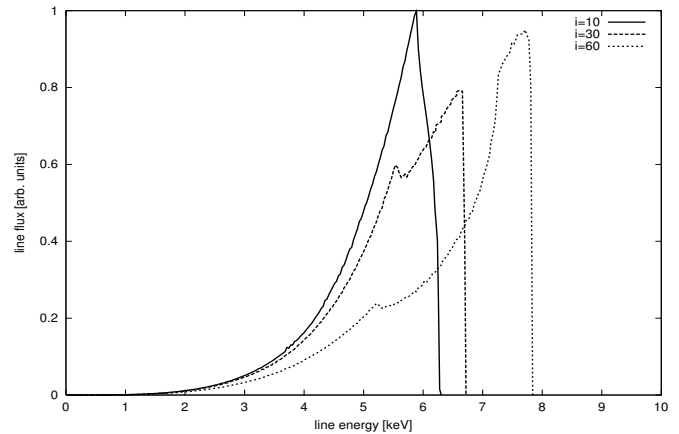}}
  \caption{Reproduction of the line set with variable inclination, $i=10^{\circ}, \ 30^{\circ}, \ 60^{\circ}$ in Schwarzschild \citep{fab}. The
  main difference is that here an additional radial drift is adapted whereas \citet{fab} only assumed Keplerian motion.
  In both cases the emissivity follows a standard single power law. For better comparison, the abscissa is scaled to the iron K$\alpha$
  line with 6.4 keV rest frame energy.} \label{fig:Fabtest}
\end{figure}
To demonstrate the \textbf{influence of radial drift}, a line set originating from a disk at three different inclination angles,
$i=10^{\circ}, \ 30^{\circ}, \ 60^{\circ}$, around a Schwarzschild black hole that extends from $r_\mathrm{in}=r_\mathrm{ms}=6.0 \ r_\mathrm{g}$ to $r_\mathrm{out}=30.0 \ r_\mathrm{g}$
with only Keplerian plasma motion is reproduced \citep{fab}. The same parameters are adapted but the inner disk edge is chosen to be at the horizon. Then,
a component $v^\mathrm{(r)}\not= 0$, e.g. a radial drift that starts at $R_\mathrm{t}=6.0 \ r_\mathrm{g}$ is considered. The emissivity law is chosen to a single power law with $\beta=3.0$
{\em without} inner cut--off to demonstrate the influence of emitting radially drifting plasma for $r\leq r_\mathrm{ms}$. Fig. \ref{fig:Fabtest} depicts the
result. A comparison of both line sets confirms that the blue edges remain almost constant. But radial drift starting already at the radius of marginal
stability, $r_\mathrm{ms}=6.0 \ r_\mathrm{g}$ in Schwarzschild, suppresses the red relic Doppler peak by enhanced gravitational redshift effects. The red wings are
in all three cases more smeared out with radial drift than in the case where only rotation is considered. For observation, this means that radial drift
produces longer and more intense red tails of the emission line.\\
Now, the relativistic emission line profiles resulting from the new \textbf{truncated accretion disk model} presented here is investigated.
The radial drift superimposed to the Keplerian motion as introduced in Sect. \ref{sec:emit} and the new emissivity model with Gaussian
shapes as has been shown in Sect. \ref{sec:emiss} are implemented.
\begin{figure}
  \rotatebox{-90}{\includegraphics[height=9cm,width=5.79cm]{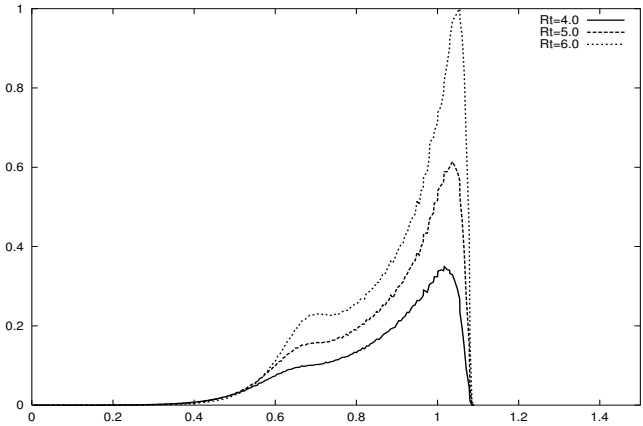}}
  \caption{Relativistic emission lines with radial drift model and Gaussian emissivity. The drift radius $R_\mathrm{t}$ from the plasma velocity field
  couples to the emissivity. From bottom to top, the drift radius increases, $R_\mathrm{t}=4.0, \ 5.0, \ 6.0$. The black hole rotates slowly, $a=0.1$, the
  inclination angle is $40^{\circ}$, $r_\mathrm{in}=r_\mathrm{H}=1.995 \ r_\mathrm{g}$ and $r_\mathrm{out}=30.0 \ r_\mathrm{g}$. The emissivity is modelled
  with constant $\sigma_\mathrm{r}=3.0$.} \label{fig:driftnGauss}
\end{figure}
The missing link to this approach is the {\em coupling} of the Gaussian emissivity to the radial drift. This model can be justified by the
following scenario: radiative accretion disks are \textbf{truncated} by efficient radiative cooling. This means that matter at the inner edge (at $R_\mathrm{t}\gtrsim r_\mathrm{ms}$)
of the standard disk evaporates and matter clouds start to fall freely into the hole. This can be modelled with the radial velocity component
$v^\mathrm{(r)}$ as introduced in Eq. (\ref{eq:vr}). At the horizon of the black hole, the plasma is mostly advected into the Kerr black hole and passes the horizon with the speed of
light (analytically derived in Eq. (\ref{eq:vr1}) and illustrated in Fig. \ref{fig:v_rvar}). The emissivity is modelled in a consistent manner:
the illumination of the inner edge of the standard disk is expected to be increased because the disk exposes a larger surface to the primary X--ray
source, e.g. a hot inner torus, an ADAF or generally speaking a corona. Therefore, the emissivity is higher than elsewhere on the accretion disk.
This can be modelled by Gaussian emissivities that decrease very steeply to both wings of the emissivity shape. This is plausible, too, because
at smaller radii there is only free--falling material that is optically thin and besides it is too hot to permit the fluorescence transition.
At larger radii, the emissivity is expected to decrease, too, because the the surface of the disk becomes too cold for fluorescence. Hence,
one can motivate to link the Gaussian profile of the emissivity to the drift radius. In a first step, \textbf{broad Gaussians} with constant
width, $\sigma_\mathrm{r}=3.0$, are applied in all three cases depicted in Fig. \ref{fig:driftnGauss}. The black hole rotation state is
constant, $a=0.1$. The drift radius varies from $R_\mathrm{t}=4.0, \ 5.0, \ 6.0 \ r_\mathrm{g}$ and is inserted in both, radial drift
(Eq. (\ref{eq:vr})) {\em and} Gaussian emissivity (Eq. (\ref{eq:gauss})). Fig. \ref{fig:driftnGauss} now shows that the line flux increases
with increasing drift radius. The calculated line width, in principle the area below the line, even grows by a factor of 2, if one compares
$R_\mathrm{t}=4.0 \ r_\mathrm{g}$ with $R_\mathrm{t}=6.0 \ r_\mathrm{g}$. This is because the Gaussian width remains constant whereas the emitting ring (limited by Gaussian
emissivity) grows in surface area. But a changing line topology can be derived, too. The red shoulder is elongated at $R_\mathrm{t}=4.0 \ r_\mathrm{g}$ because of
enforced gravitational redshift for smaller drift radii respectively truncation closer to the black hole. Besides, the simulation yields a slight
variation in the line form. This can be proven with the quantity $\mathrm{DPR}$, the \textbf{Doppler Peak Ratio} which is independent from absolute flux values.
This is the ratio of the flux from the blue Doppler peak versus the red Doppler peak. Surely, this can only be calculated if the two Doppler peaks
exist. These $\mathrm{DPRs}$ are $\mathrm{DPR}(R_\mathrm{t}=4.0)=3.6$, $\mathrm{DPR}(R_\mathrm{t}=5.0)=4.0$ and $\mathrm{DPR}(R_\mathrm{t}=6.0)=4.3$. So, the red shoulder drops down with increasing $R_\mathrm{t}$ and
the blue flux peak increases by the growth of the emitting region.\\
An alternative to the above issue, is to replace the Gaussian emissivity by the \textbf{cut--emissivity profile} as introduced in Eq. (\ref{eq:cut}).
This radial shape ressembles to the coronal emissivity profiles recently presented \citep{MF}. For comparative reasons, all other parameters as
in the example with Gaussian emissivity are maintained and only the emissivity is changed to a cut--power law with constant $\alpha=3.0$ and the
same truncation radii set, $R_\mathrm{t}=4.0, \ 5.0, \ 6.0$. The behaviour at larger radii corresponds to the standard single power law $\propto r^{-3.0}$. The
inferred emissivities are shown in Fig. \ref{fig:cutem456}.
The resulting line profiles with these cut--emissivity shapes can be investigated in Fig. \ref{fig:driftncpl}.
For the three lines itself, the allover topologies are almost identical, even $\mathrm{DPS}$ remains constant. But the ratio of the relic Doppler peaks slightly
increases with $R_\mathrm{t}$: $\mathrm{DPR}(R_\mathrm{t}=4.0)=4.95$, $\mathrm{DPR}(R_\mathrm{t}=5.0)=5.16$ and $\mathrm{DPR}(R_\mathrm{t}=6.0)=5.26$. This effect is mainly caused by the fall of the red
wing driven by the radial drift. Remarkable is the fact that the blue edge stays absolutely constant. Comparing Figs. \ref{fig:driftnGauss} and
\ref{fig:driftncpl}, the blue edges remain constant. But, it is stressed that at the parameters chosen the Gaussian emissivity produces {\em shoulder--like}
line shapes whereas the cut--power law emissivity produces {\em double--horned} profiles. \\
The reduction of the red wing flux caused by including radial drift is a generic feature. Fig. \ref{fig:Kepvsdrift} confirms this. Here two lines are compared
where in one case a pure Keplerian velocity field applies and in the second case the radial drift model is used. This study reveals immediately the influence
of including plasma drift. In this example, the drift starts at $R_\mathrm{t}=4.5$. One can see that the red part of the line drops down when drift is operating.
The interpretation is that the gravitational redshift is enhanced by the drift, compare Eq. (\ref{eq:g}), and results in a lower red wing flux. This statement holds in principle for all
simulated lines where the drift model is compared to its pure Keplerian counterpart.\\
Finally, a more complicated line profile with a few species in application to a recent observation of Seyfert--1 galaxy
MCG--6--30--15 in the high state \citep{fab2} is demonstrated. The characteristic shoulder--like shape of this observation can be reproduced theoretically with
the following plausible parameters:
\begin{figure}
  \rotatebox{0}{\includegraphics[height=4.94cm,width=9cm]{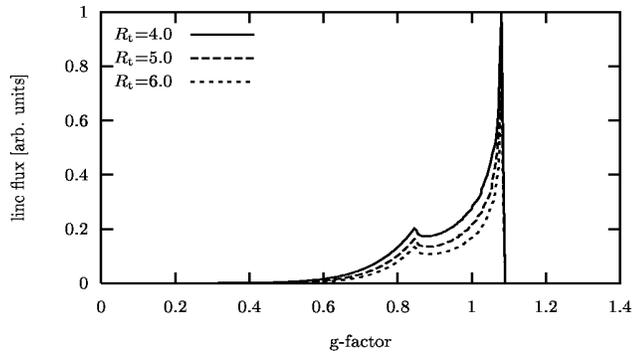}}
  \caption{Relativistic emission lines with radial drift model and cut--power law emissivity. The drift radius $R_\mathrm{t}$ from the plasma velocity field
  couples to the emissivity. From bottom to top, the drift radius increases, $R_\mathrm{t}=4.0, \ 5.0, \ 6.0$. The black hole rotates slowly, $a=0.1$, the
  inclination angle is $40^{\circ}$, $r_\mathrm{in}=r_\mathrm{H}=1.995 \ r_\mathrm{g}$ and $r_\mathrm{out}=30.0 \ r_\mathrm{g}$. The emissivity is modelled with constant $\alpha=\beta=3.0$.} \label{fig:driftncpl}
\end{figure}
\begin{itemize}
\item rapidly spinning black hole, $a=0.8$,
\item typical Seyfert--1 inclination, $i=40^{\circ}$,
\item inner edge at horizon, $r_\mathrm{in}=r_\mathrm{H}=1.6 \ r_\mathrm{g}$,
\item outer edge at $r_\mathrm{out}=20.0 \ r_\mathrm{g}$,
\item standard Keplerian rotation \textbf{plus}
\item radial drift starting at $R_\mathrm{t}=4.5 \ r_\mathrm{g}$,
\item Gaussian emissivity: $R_\mathrm{t}=4.5 \ r_\mathrm{g}$ and $\sigma_\mathrm{r}=0.4 \ R_\mathrm{t}$,
\item multi--species line system consisting in Fe K$\alpha$, Fe K$\beta$, Ni K$\alpha$ and Cr K$\alpha$
\end{itemize}
The relative line strength of the different species are modelled in an approximation found in Reynolds (1996). Fig. \ref{fig:Fabfit} now
illustrates a line system consisting of iron, nickel and chromium. The total observed X--ray fluorescence feature around 7 keV is equal
to the sum of these contributions. The theoretically inferred line FWHM is very high, around 1 keV, which has the same order as the
$\approx 750$ eV as can be found in \citet{fab2}. These parameters fit nicely the observed shoulder--like red wing
and even the small blue bump beyond the Fe K$\alpha$ beaming branch. The flux ratio between shoulder-level and maximal flux around a factor of 5 is close
to the observed value of 4. The other elements seem to be vital to model this observed profile. However, chromium seems to play a minor role due to weak
relative strength.
%
%
%
%
\section{Conclusions}
It has been demonstrated that radial drift, a non--vanishing radial velocity component, is a vital ingredient in modelling relativistic emission line
profiles. Accretion theory demands on such more complicated non--Keplerian velocity fields. However, our approach is just a phenomenological model
motivated by simulations of accreting black hole systems. Some analytical features, as the asymptotical behaviour at the black hole horizon are
independent of these models. One main consequence is, that the radial ZAMO velocity becomes the speed of light at the horizon whereas the toroidal
ZAMO velocity vanishes there. Another essential feature in accreting and radiating black hole systems is the steep decrease of the generalized Doppler
factor causing a shadow of the black hole. Hence, any emission near black holes is strongly suppressed. This is the competing dominant effect that
suppresses as well steep emissivity profiles within the marginally stable orbit. \\
The truncated standard disk (TSD) hypothesis induced by radiative pseudo--Newtonian hydrodynamics is a new aspect in accretion theory that also
influences the models for X--ray spectroscopy, especially the formation of broad relativistic emission lines. Future fully relativistic {\em and}
radiative MHD is expected to confirm this. Then, a coupling of 3D ray tracing techniques and plasma dynamics are inevitable to explain the morphology
and variablity of relativistic emission lines, particularly the iron K$\alpha$ line. Besides, one hope is that these demanding simulations will
enlighten the open question of the corona topology. The answer is essential to understand X--ray irradiation of thin accretion disks. \\
Alternative disk emissivity models, such as cut--power laws and Gaussian emissivities can serve as valuable tools to study the huge parameter space of
the line zoo. The localized emissivity studies with Gaussian emissivity laws prove that emission from regions neighboring the black hole horizon is
massively suppressed by the low generalized Doppler factor. Additionally, the exponentially decreasing cut--power law may be a suitable profile function
to model the disk emissivities with higher magnetically driven dissipation at the inner disk edge. \\
A first classification of the variety of relativistically broadened emission lines succeeds by a topological criterion: the lines can be divided in
families with triangular, double--horned, double--peaked, bumpy and shoulder-like line form. It turns out that this first step constrains the parameter
space significantly.\\
Unfortunately, it is still not feasible to solve the fully covariant radiation transport problem. Therefore, one deals always with some
restrictions in the accretion models: either the approach is fully relativistic, but neglects radiation or the model incorporates radiation
processes, but mimics relativity only to a certain degree by pseudo--Newtonian potentials. The future of accretion theory points towards
\textbf{radiative 3D--GRMHD}. Until this will be possible, one has to extract few informations from 'simpler' models with
limitations done in the past (such as pseudo--Newtonian models , non--radiative issues, less dimensions etc.) and puzzle this
input to an understanding of black hole accretion physics. \\
%
%
%
%

\begin{acknowledgements}
Part of this work was supported by the {\em Deutsche Forschungsgemeinschaft, DFG} (Sonderforschungsbereich 439).
\end{acknowledgements}

\bibliographystyle{aa}
\bibliography{RelEmLines09-2003}

\begin{thebibliography}{}

\bibitem[\protect\astroncite{Abramowicz et al.}{1978}]{abram}
Abramowicz, M., Jaroszynski, M., Sikora, M. 1978,
\newblock {A\&A}, {63}, 221

\bibitem[\protect\astroncite{Armitage \& Reynolds}{2003}]{AR}
Armitage, P.J., Reynolds, C.S. 2003,
\newblock {astro-ph/} {0302271}, submitted to MNRAS

\bibitem[\protect\astroncite{Balbus \& Hawley}{1991}]{BH2}
Balbus, S.A., Hawley, J.F. 1991,
\newblock {ApJ}, {376}, 214

\bibitem[\protect\astroncite{Balbus \& Hawley}{2002}]{BH}
Balbus, S.A., Hawley, J.F.  2002,
\newblock {ApJ}, {573}, 738

\bibitem[\protect\astroncite{Ballantyne et al.}{2001}]{BRF}
Ballantyne, D.R., Ross, R.R., Fabian, A.C. 2001,
\newblock {MNRAS}, {327}, 10

\bibitem[\protect\astroncite{Blandford \& Znajek}{2000}]{bz}
Blandford, R.D., Znajek, R.L. 1977,
\newblock {MNRAS}, {179}, 433

\bibitem[\protect\astroncite{Bromley et al.}{1997}]{brom}
Bromley, B.C., Chen, K., Miller, W.A. 1997,
\newblock {ApJ}, {475}, 57

\bibitem[\protect\astroncite{Carter}{1968}]{cart}
Carter, B. 1968,
\newblock {Phys. Rev.}, {174}, 1559

\bibitem[\protect\astroncite{Chandrasekhar}{1983}]{chan}
Chandrasekhar, S. 1983,
\newblock {The Mathematical Theory of Black Holes} {(Oxford University Press, 1983)}

\bibitem[\protect\astroncite{Cunningham}{1975}]{cunn}
Cunningham, C.T. 1975,
\newblock {ApJ}, {202}, 788

\bibitem[\protect\astroncite{Dabrowski \& Lasenby}{2001}]{dabr}
Dabrowski, Y., Lasenby, A.N. 2001,
\newblock {MNRAS}, {321}, 605

\bibitem[\protect\astroncite{De Villiers \& Hawley}{2003}]{DeVH}
De Villiers, J.P., Hawley, J.F.  2003,
\newblock {ApJ}, {592}, 1060

\bibitem[\protect\astroncite{Dewangan et al.}{2003}]{DGS}
Dewangan, G.C., Griffiths, R.E., Schurch, N.J. 2003,
\newblock {ApJ}, {592}, 52

\bibitem[\protect\astroncite{Esin et al.}{1997}]{esin}
Esin, A.A., McClintock, J.E., Narayan, R. 1997,
\newblock {ApJ}, {489}, 865

\bibitem[\protect\astroncite{Fabian et al.}{1989}]{fab3}
Fabian, A.C., Rees, M.J., Stella, L., White, N.E. 1989,
\newblock {MNRAS}, {238}, 729

\bibitem[\protect\astroncite{Fabian}{1998}]{fab4}
Fabian, A.C. 1998,
\newblock {Astronomy \& Geophysics}, {123}

\bibitem[\protect\astroncite{Fabian et al.}{2000}]{fab}
Fabian, A.C., Iwasawa, K., Reynolds, C.S., Young, A.J. 2000,
\newblock {PASP}, {112}, 1145

\bibitem[\protect\astroncite{Fabian et al.}{2002}]{fab2}
Fabian, A.C., Vaughan, S., Nandra, K., et al. 2002,
\newblock {MNRAS}, {335}, L1

\bibitem[\protect\astroncite{Falcke et al.}{2000}]{fal}
Falcke, H., Melia, F., Agol, E. 2000,
\newblock {ApJ}, {528}, L13

\bibitem[\protect\astroncite{Fanton et al.}{1997}]{fant}
Fanton, C., Calvani, M., Felice, F.de, Cadez, A. 1997,
\newblock {PASJ}, {49}, 159

\bibitem[\protect\astroncite{Gracia}{2002}]{jose}
Gracia-Calvo, J. 2002,
\newblock {Time-dependent accretion flows onto black holes, Ph.D. thesis,} {Landessternwarte Heidelberg}

\bibitem[\protect\astroncite{Gracia}{2003}]{josep}
Gracia-Calvo, J. 2003,
\newblock {astro-ph/} {0301113}, submitted to MNRAS

\bibitem[\protect\astroncite{Hujeirat \& Camenzind}{2000}]{huj}
Hujeirat, A., Camenzind, M. 2000,
\newblock {A\&A}, {361}, L53

\bibitem[\protect\astroncite{Iwasawa et al.}{1996}]{iwa}
Iwasawa, K., 1996,
\newblock {MNRAS}, {282}, 1038

\bibitem[\protect\astroncite{Kerr}{1963}]{kerr}
Kerr, R.P. 1963,
\newblock {Phys. Rev. Lett.}, {11}, 237

\bibitem[\protect\astroncite{Lee}{2001}]{lee}
Lee, J.C., Ogle, P.M., Canizares, C.R., et al. 2001,
\newblock {ApJ}, {554}, L13

\bibitem[\protect\astroncite{Manmoto \& Kato}{2000}]{man}
Manmoto, T., Kato, S. 2000,
\newblock {ApJ}, {538}, 295

\bibitem[\protect\astroncite{Mason et al.}{2003}]{mas}
Mason, K.O., Branduardi-Raymont, G., Ogle, P.M., et al. 2003,
\newblock {ApJ}, {582}, 95

\bibitem[\protect\astroncite{Matt et al.}{1993}]{matt}
Matt, G., Fabian, A.C., Ross, R.R. 1993,
\newblock {MNRAS}, {262}, 179

\bibitem[\protect\astroncite{Matt et al.}{1997}]{matt2}
Matt, G., Fabian, A.C., Reynolds, C.S. 1997,
\newblock {MNRAS}, {281}, 175

\bibitem[\protect\astroncite{Merloni \& Fabian}{2003}]{MF}
Merloni, A., Fabian, A.C. 2003,
\newblock {MNRAS}, {342}, 951

\bibitem[\protect\astroncite{M\"uller}{2000}]{andy}
M\"uller, A. 2000,
\newblock {Emissionslinienprofile akkretierender Scheiben um rotierende Schwarze L\"ocher, diploma thesis,} {Landessternwarte Heidelberg}

\bibitem[\protect\astroncite{Narayan \& Yi}{1994}]{naryi}
Narayan, R., Yi, I. 1994,
\newblock {ApJ}, {428}, L13

\bibitem[\protect\astroncite{Novikov \& Thorne}{1974}]{NT}
Novikov, I.D., Thorne, K.S. 1974,
\newblock {Black Holes}, {343}

\bibitem[\protect\astroncite{Page \& Thorne}{1974}]{PT}
Page, D.N., Thorne, K.S. 1974,
\newblock {ApJ}, {499}, 191

\bibitem[\protect\astroncite{Penrose \& Floyd}{1971}]{PF}
Penrose, R., Floyd, R.M. 1971,
\newblock {Nature Phys. Sci.}, {229}, 177

\bibitem[\protect\astroncite{Pfefferkorn et al.}{2001}]{pfeff}
Pfefferkorn, F., Boller, T., Rafanelli, P. 2001,
\newblock {A\&A}, {368}, 797

\bibitem[\protect\astroncite{Pounds et al.}{1990}]{poun}
Pounds, K.A., Nandra, K., Stewart, G.C., George, I.M., Fabian, A.C., et al. 1990,
\newblock {Nature}, {344}, 132

\bibitem[\protect\astroncite{Reeves et al.}{2001}]{boll}
Reeves, J.N., Turner, M.J.L., Pounds, K.A., et al. 2001,
\newblock {A\&A}, {365}, L134

\bibitem[\protect\astroncite{Reynolds}{1996}]{reyn}
Reynolds, C.S. 1996,
\newblock {X-ray emission and absorption in active galactic nuclei, Ph.D. thesis,} {University of Cambridge}

\bibitem[\protect\astroncite{Reynolds \& Nowak}{2003}]{RN}
Reynolds, C.S., Nowak, M.A. 2003,
\newblock {Physics Rep.}, {377}, 389

\bibitem[\protect\astroncite{Ross et al.}{1999}]{RFY}
Ross, R.R., Fabian, A.C., Young, A.J. 1999,
\newblock {MNRAS}, {306}, 461

\bibitem[\protect\astroncite{Rybicki \& Lightman}{1979}]{RL}
Rybicki, G.B., Lightman, A.P. 1979,
\newblock {Radiative Processes in Astrophysics,} {John Wiley \& Sons, New York, 1979}

\bibitem[\protect\astroncite{Sch\"odel et al.}{2002}]{schoe}
Sch\"odel, R., Ott, T., Genzel, R., et al. 2002,
\newblock {Nature}, {419}, 694

\bibitem[\protect\astroncite{Shakura \& Sunyaev}{1973}]{SS}
Shakura, N.L., Sunyaev, R.A. 1973,
\newblock {A\&A}, {24}, 337

\bibitem[\protect\astroncite{Speith et al.}{1995}]{spei}
Speith, R., Riffert, H., Ruder, H. 1995,
\newblock {Comput. Phys. Commun.}, {88}, 109

\bibitem[\protect\astroncite{Spindeldreher}{2002}]{spindel}
Spindeldreher, S. 2002,
\newblock {The Discontinuous Galerkin Method applied on the equations of ideal relativistic hydrodynamics, Ph.D. thesis,} {Landessternwarte Heidelberg}

\bibitem[\protect\astroncite{Tanaka et al.}{1995}]{tan}
Tanaka, Y., Nandra, K., Fabian, A.C., et al. 1995,
\newblock {Nature}, {375}, 659

\bibitem[\protect\astroncite{Turner et al.}{1997}]{turn}
Turner, T.J., George, I. M., Nandra, K., Mushotzky, R. F. 1997,
\newblock {ApJ}, {488}, 164

\bibitem[\protect\astroncite{Wang et al.}{1999}]{wang}
Wang, J.-X., Zhou, Y.-Y., Wang, T.G. 1999,
\newblock {MNRAS}, {328}, L27

\bibitem[\protect\astroncite{Wilms et al.}{2001}]{wilms}
Wilms, J., Reynolds, C.S., Begelman, M.C., et al. 2001,
\newblock {MNRAS}, {328}, L27

\bibitem[\protect\astroncite{Yaqoob et al.}{1999}]{yaq}
Yaqoob, T., et al. 1999,
\newblock {ApJ}, {525}, L9
\end{thebibliography}

\end{document}